\documentclass[a4paper,usenatbib]{mnras}
\usepackage[utf8]{inputenc}
\usepackage{xspace}
\usepackage{amsmath}
\usepackage{amssymb}
\usepackage{epsfig}
\usepackage{defs}

\title[Missing Baryons and Core Creation ]{Stellar Feedback and the Energy Budget of Late-Type Galaxies: Missing Baryons and Core Creation }

\author[H. Katz et al. ]{Harley Katz$^{1}$\thanks{E-mail:harley.katz@physics.ox.ac.uk}, Harry Desmond$^1$\thanks{St. John's JRF}, Federico Lelli$^2$\thanks{ESO Fellow}, Stacy McGaugh$^3$, \newauthor Arianna Di Cintio$^{4,5}$, Chris Brook$^{4,5}$ and James Schombert$^6$\\
$^1$Astrophysics, University of Oxford, Denys Wilkinson Building, Keble Road, Oxford OX1 3RH, UK\\
$^2$European Southern Observatory, Karl-Schwarzschild-Strasse 2, Garching bei Munchen, Germany\\
$^3$Department of Astronomy, Case Western Reserve University, Cleveland, OH 44106, USA\\
$^4$Instituto de Astrof\'{i}sica de Canarias, Calle Via L\'{a}ctea s/n, E-38206 La Laguna, Tenerife, Spain\\
$^5$Universidad de La Laguna. Avda. Astrof\'{i}sico Fco. S\'{a}nchez, La Laguna, Tenerife, Spain\\
$^6$University of Oregon, Department of Physics, Eugene OR, 97403, USA\\
} 

\begin{document}

\maketitle
\begin{abstract}
In a $\Lambda$CDM cosmology, galaxy formation is a globally inefficient process: it is often the case that far fewer baryons are observed in galaxy disks than expected from the cosmic baryon fraction. The location of these ``missing baryons'' is unclear. By fitting halo profiles to the rotation curves of galaxies in the SPARC data set, we measure the ``missing baryon'' mass for individual late-type systems. Assuming that haloes initially accrete the cosmological baryon fraction, we show that the maximum energy available from supernovae is typically not enough to completely eject these ``missing baryons'' from a halo, but it is often sufficient to heat them to the virial temperature. The energy available from supernovae has the same scaling with galaxy mass as the energy needed to heat or eject the ``missing baryons'', indicating that the coupling efficiency of the feedback to the ISM may be constant with galaxy virial mass. We further find that the energy available from supernova feedback is always enough to convert a primordial cusp into a core and has magnitude consistent with what is required to heat the ``missing baryons'' to the virial temperature. Taking a census of the baryon content of galaxies with ${\rm 10^9<M_{vir}/M_{\odot}<10^{12}}$ reveals that $\sim86\%$ of baryons are likely to be in a hot phase surrounding the galaxies and possibly observable in the X-ray, $\sim7\%$ are in the form of cold gas, and $\sim7\%$ are in stars.
\end{abstract}

\begin{keywords}
galaxies: general, galaxies: formation, galaxies: evolution, galaxies: haloes, galaxies: spiral, galaxies: fundamental parameters
\end{keywords}

\section{Introduction}
In the early Universe, it is predicted that dark matter and baryons are well mixed \citep{Spergel2003}.  As dark matter haloes form by gravitational collapse, the baryons cool, dissipate energy, and fall to the centres of the haloes \citep{White1978,Fall1980}.  Assuming a simple model where no other processes are at play, the total mass of a galaxy should then be composed of $\sim15\%$ baryons and $\sim85\%$ dark matter \citep{Planck2015}.  However, only in the most massive galaxy clusters does the observed baryon fraction begin to approach this value \citep{Giodini2009,Gonzalez2013}, while at dwarf galaxy masses, only $\sim1\%$ of the expected baryon content is actually detected \citep{McGaugh2010,Papastergis2013,Bradford2015}. A simple observational census of the baryonic content of the Universe therefore reveals that most baryons have not been identified.

Many of these ``missing baryons" may reside in the inter-galactic medium (IGM), although this is unlikely to account for the full expected quantity  \citep{Fukugita1998,Danforth2005,Shull2012}.  Alternatively, these ``missing baryons" may exist in a phase that is difficult to detect \citep{Bregman2015}.  Accretion onto dark matter haloes can shock heat the gas up to the virial temperature (${\rm T_{vir}}$), which would make the gas emit primarily in the X-ray \citep{White1991,Cen1999}.  Many hydrodynamic simulations show that galaxies, especially at high redshift, are primarily fed by cold flows which penetrate deep into the centres of the haloes \citep{Keres2005}.  However, this may not be the case at low redshift. Strong feedback processes resulting from galaxy formation can reheat much of this gas into a hot phase, placing the primary emission in the X-ray \citep{Mathews1971,McKee1977,Cen1999}. Hot gaseous haloes have been detected around a number of galaxies \citep{Anderson2011,Anderson2013,Miller2015,Bregman2018}, although for massive spirals, this may not completely account for the entire budget of ``missing baryons" \citep{Li2018,Bregman2018}. Hot baryons may also be detected by the thermal Sunyaev-Zel'dovitch (kSZ) effect \citep{tSZ1, tSZ2}.  As observations improve more of the ``missing baryons" are being found, but it is unclear whether they are sufficient to make up the baryon content of the Universe.

This inefficiency by which galaxies either obtain or retain their baryons translates directly into the stellar content of galaxies.  Abundance matching techniques predict that even for the most efficient star-forming galaxies, the fraction of stars present  remains far below the cosmic baryon fraction at all masses \citep{Moster2013, Behroozi2013, Moster2017}.  Explanations for the low fraction of stars come from feedback in three regimes.  At the very lowest halo masses, (${\rm M_{halo}\lesssim10^9\ M_{\odot}}$), reionization may have completely prevented galaxies from forming by limiting their ability to accrete gas \citep{Babul1992,Efstathiou1992,Gnedin2000,Okamoto2008}.  At slightly higher masses, supernova (SN) feedback can limit star formation by either ejecting gas from the halo or keeping it at a temperature ${\rm T \sim T_{vir}}$ \citep{Dekel1986}.  Finally, at the highest mass systems, feedback from AGN may be the dominant mechanism which regulates galaxies \citep{Silk1998,Bower2006}.

Feedback processes can impact galaxies in many different ways. For example, the slope of the predicted black hole mass--stellar velocity dispersion relation depends on whether AGN outflows are ``momentum-driven" or ``energy-driven" \citep{King2003,Costa2014}.  For stellar feedback, the impulsiveness of the energy injection can determine whether or not a dark matter cusp can be converted into a core \citep{Navarro1996b,Pontzen2012}.  The way in which the feedback is modelled has drastic effects on observable quantities such as the stellar mass function and the galaxy distribution along the Hubble sequence \citep[e.g.][]{Schaye2015,Vogelsberger2014,Dubois2014,Hopkins2017}, bulge formation \citep[e.g.][]{Hopkins2012}, central black hole mass \citep[e.g.][]{Curtis2016,DiCintio2017}, gas fractions and X-ray luminosities \citep[e.g.][]{Puchwein2008}. Understanding the energy scales involved and how the feedback energy couples to the local medium is key to developing a complete model for galaxy formation.

Historically, one of the most well studied effects of feedback is how the density profile of the dark matter halo responds to large scale inflows or outflows of baryons \citep{Navarro1996,Pontzen2012,Blumenthal1986,Gnedin2004,Katz2014,Martizzi2013,Read2005,Read2016}. \cite{Penarrubia2012} calculated the range of energies required to transform a cuspy density profile to one with a core and compared it to the energy budget of SN.  They identified that Milky Way dSphs require $10^{53-55}$~ergs of energy in order to form cores with sizes comparable to the luminous size of the galaxies.  Using similar analytic reasoning, we focus here on how stellar feedback impacts the slope of the ${\rm M_*-M_{halo}}$ relation for late-type galaxies (spirals and dIrr) as well as dark matter density profiles.

We focus on two halo profiles: NFW \citep{Navarro1996}, which is a prediction from cosmological DM-only simulations, and DC14 \citep{DC2014,DiCintio2014b}, which is a fit to the halos produced in the MaGICC simulations \citep{Brook2012,Stinson} that include the effects of galaxy formation. This model includes a dependence of halo shape on ${\rm M_*/M_{halo}}$ such that the dark matter profiles of low and high mass galaxies are cuspy at the centre, while intermediate mass galaxies exhibit cores. Although this model is successful at accounting for the kinematics of a range of galaxy types \citep{DC2014,Katz2016} it is important to bear in mind that the formation of cores is a contentious issue among the current-generation of cosmological hydrodynamic simulations. For example, although other high-resolution simulations produce cores of roughly the DC14 type \citep[e.g.][]{Chan, Read}, others do not produce cores at all \citep[e.g.][]{Apostle}.  The exact mechanism of core formation has yet to be determined and there are indeed degeneracies in subgrid implementations of star formation and feedback that may lead to the same results.  Nevertheless, since the DC14 model fits the properties of observed rotation curves well, we aim to use the energy scales of real galaxies as a further test of whether SN can provide enough energy to both regulate star formation and produce cores in real galaxies. 

We begin by presenting observational results on the fraction of cold gas and stars of $\sim150$ late-type galaxies from the SPARC data set \citep{Lelli2016} (Sec.~\ref{obscon}).  We then review the theoretical motivation for how SN feedback regulates galaxy formation and show that in two simple models it naturally leads to a logarithmic slope of $5/3$ for the ${\rm M_*-M_\text{vir}}$ relation (Sec.~\ref{theory}).  These models are then tested against the SPARC data where this slope can be independently measured: we then check whether this energy budget is sufficient to generate a core in low mass galaxies or reverse adiabatic contraction in higher mass systems (Sec.~\ref{comparison}).  Finally, in Sec.~\ref{discussion} we use these results to take a cosmic census of the distribution of baryons in late-type galaxies with $10^9<{\rm M_{vir}}<10^{12}$.  Throughout this work, we assume a WMAP3 cosmology with $H_0 = 73{\rm km~s^{-1}~Mpc^{-1}}$, $\Omega_{\rm m} = 0.24$, $\Omega_{\Lambda}= 0.76$, $\Omega_{\rm b} = 0.04$ and $\sigma_8 = 0.76$ \citep{Spergel2007}, and we define the virial radius to be the radius which contains a mean density equal to $93.6\:\rho_{\rm crit}$.

\section{Observational Constraints}
\label{obscon}
\subsection{SPARC Dataset and Rotation Curve Fitting}
The SPARC sample\footnote{\url{astroweb.cwru.edu/SPARC/}} contains 175 galaxies with high quality rotation curves, interferometric observations at 21 cm, and baryon mass models from \textit{Spitzer} [3.6 $\mu$m] photometry. We use this data to measure $f_d \equiv {\rm (M_{gas,cold}+M_*)/([\Omega_b/\Omega_{m}]M_{vir})}$, the fraction of the galaxy mass that has settled into a cold disk, relative to the cosmological value, and $f_* \equiv {\rm M_*/(M_{gas,cold}+M_*)}$, the fraction of baryons in stars. SPARC is ideal for this purpose as it contains late-type galaxies with a wide range of properties: $3\times10^7\lesssim {\rm M_*}\ {\rm (M_{\odot})}\lesssim3\times10^{11}$, $20\lesssim {\rm V_c\ (km/s)}\lesssim300$, and $3\lesssim\Sigma_*\ {\rm (M_{\odot}/pc^2)}\lesssim1500$.  We extract from the SPARC data set a subsample of 147 galaxies which pass a series of selection criteria: 1) $i \ge 30^o$, 2) at least 5 observed rotation curve points, and 3) no major kinematic asymmetries between the approaching and receding sides of the disc (quality flag $Q<3$).

We follow \cite{Katz2016} in fitting the rotation curves with either NFW or DC14 profiles. We fit these profiles using an MCMC algorithm with three free parameters: ${\rm V_{vir}}$, $c_{\rm vir}$ and ${\rm M_*/L}$ where $L$ is measured at $3.6\mu$m.  We define a set of ``flat" priors such that $10<{\rm V_{vir}}\ {\rm (km/s)} < 500$, $1<c_{\rm vir}<100$, and $0.3<{\rm M_*/L}<0.8$.  The first two priors ensure that no walkers end up in extremely unphysical regimes. The prior on ${\rm M_*/L}$ is set to be consistent with \cite{Mcgaugh2014} who demonstrated that ${\rm M_*/L}$ shows no significant colour trends. \cite{Katz2016} demonstrated that the fitting results are rather robust to a variety of choices on ${\rm M_*/L}$. Note that \cite{Katz2016} also fit each galaxy while assuming an additional set of ``$\Lambda$CDM" priors, where both the ${\rm M_{halo}}-c$ and ${\rm M_*/M_{halo}}$ relations were imposed as lognormal priors on the fits. We do not include these priors here as we wish to constrain $f_d$ and $f_*$ from the data alone.

From the observations, we obtain the gas velocity, ${\rm V_{cold,gas}}(r)$, stellar velocity (dependent on ${\rm M_*/L}$), ${\rm V_{*}}(r)$, and the circular velocity, ${\rm V_c}(r)$. Hence
\begin{equation}
    {\rm V_c}^2(r)={\rm V_{cold,gas}}^2(r)+({\rm M_*/L}){\rm V_{*}}^2(r)+{\rm V_{dark}}^2(r).
\end{equation}
By fitting the rotation curves we constrain ${\rm V_{dark}}^2(r)={\rm V_{hot,gas}}^2(r)+{\rm V_{halo}}^2(r)$. ${\rm V_{hot,gas}}^2(r)=0$ corresponds to a feedback model where we assume that all unseen gas has been ejected from the halo. Alternatively, if we assume that the gas is heated to $T_\text{vir}$ then it still resides in the halo and ${\rm V_{hot,gas}}^2(r)\ne0$. Given
\begin{equation}
    {\rm M_{vir}=M_{halo}+M_{*}+M_{gas,cold}+M_{gas,hot}},
\end{equation}
\begin{equation}
    {\rm M_{dark}=M_{halo}+M_{gas,hot}},
\end{equation}
and
\begin{equation}
    {\rm M_{halo}=\frac{\Omega_{DM}}{\Omega_m}M_{vir}}
\end{equation}
we can separate the mass in hot gas from the dark matter component. We assume that the distribution of hot gas follows the same density profile as the dark matter.  This is unlikely to be completely true as the hot gas probably has a higher concentration than the dark matter, although it is doubtful that introducing this additional freedom into the fitting would be fruitful.  The difference between the models is subtle since ${\rm M_{gas,hot}}$ is at most $\sim20\%$ of the halo mass.  

\begin{figure*}
\centerline{\includegraphics[scale=1]{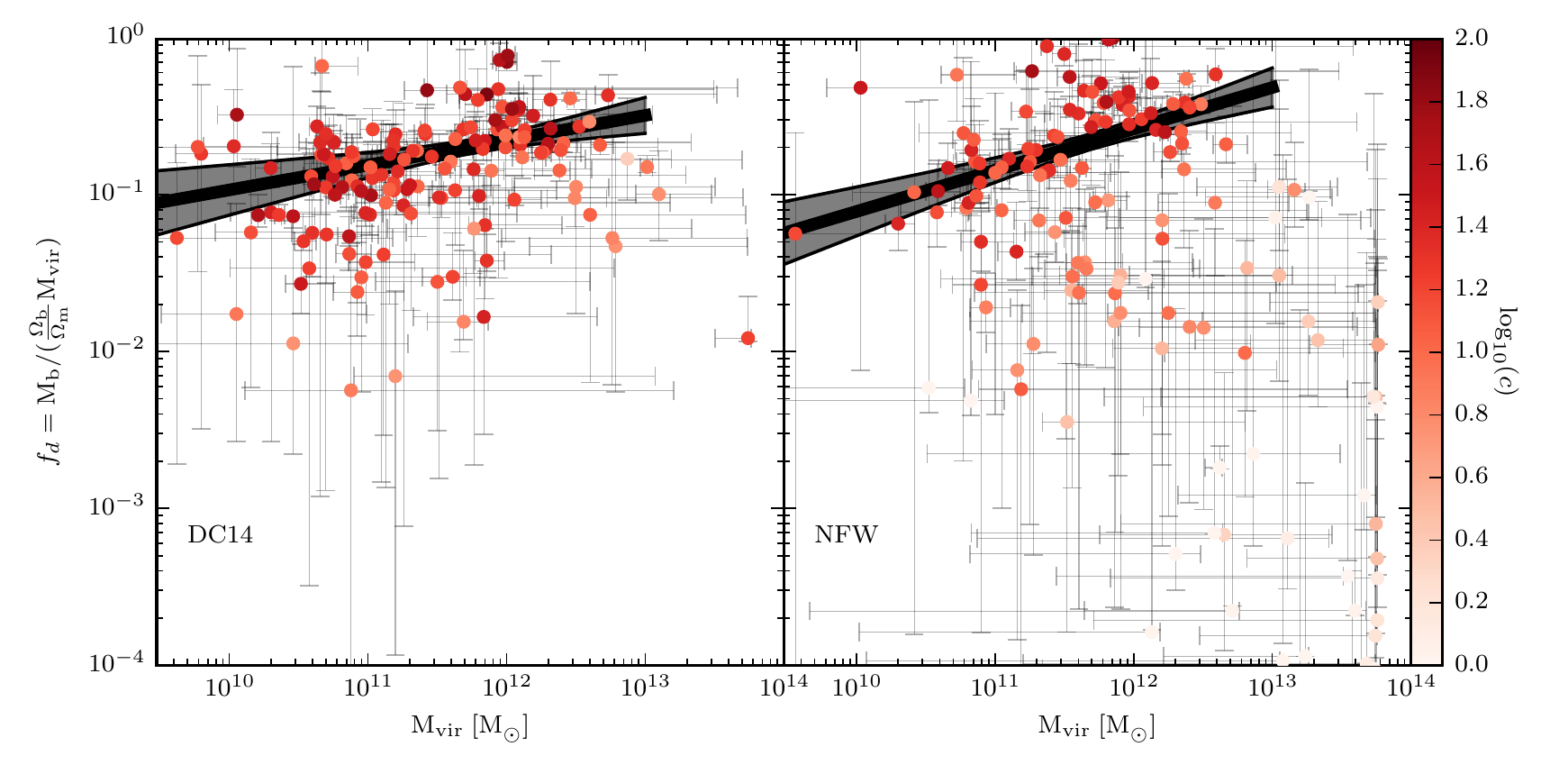}}
\caption{$f_d$ as a function of the virial mass of the galaxy for the DC14 model (left) and the NFW model (right).  The points represent the maximum likelihood fits to the observed rotation curves and they are coloured by the log of the concentration.  The error bars represent the $2\sigma$ uncertainties on the parameters for individual galaxies.  Note that the concentrations measured for the DC14 model are in much better agreement with the expected mass-concentration relation than for the NFW model \protect\cite{Katz2016}. The thick black line shows the best fit values to the function given in Equation~\ref{fdmvir} when the RANSAC algorithm is used, and the grey shaded region shows the $1\sigma$ uncertainty on this fit.}
\label{fd}
\end{figure*}

\subsection{Estimating $f_d$ and $f_*$}
With the rotation curve fits in hand, we can now derive $f_d$ and $f_*$ directly from the observational data.  In Figure~\ref{fd}, we plot $f_d$ as a function of ${\rm M_{vir}}$  We fit these galaxies\footnote{For all fits in this paper, we only consider galaxies with virial mass $<10^{13}$M$_{\odot}$. We find similar results when fitting when using a maximum mass of $<10^{12}$M$_{\odot}$.} with a linear model in log-space such that\footnote{The $p$-value of Pearson's test on the $f_d-M_\text{vir}$ correlation for the DC14 model is 0.02, indicating a statistically significant linear relation.}
\begin{equation}
    \log_{10}(f_d)=A\log_{10}({\rm M_{vir}})+B.
    \label{fdmvir}
\end{equation}

\noindent In order to make these fits, we take into account the uncertainties on both ${\rm M_{dark}}$ and ${\rm M_*/L}$ that come from our MCMC fits.  Unfortunately, the confidence intervals on our maximum posterior fits are both asymmetric and non-gaussian so we cannot perform an ordinary least square fit.  Instead we adopt a bootstrap approach, as follows. First, we randomly choose a point in the 3D parameter space from 100,000 steps taken by the 100 walkers used in our MCMC chains, for each galaxy. We fit this resampled catalog with a RANSAC linear fitting algorithm \citep{Fischler1981,sklearn} in order to deal with potential outliers\footnote{The RANSAC algorithm is completely different from sigma clipping and uses an iterative method to robustly search the data set for inliers and outliers.}. We repeat this procedure 10,000 times. The relation we quote is given by the mean of the fits to these catalogs, and the uncertainty in the fits is given by the standard deviation of the fitting parameters.  Note that we only consider galaxies with ${\rm M_{vir}}<10^{13}$M$_{\odot}$ in our fits. 

In Table~\ref{tabfd}, we list the fitting parameters measured from the data for Equation~\ref{fdmvir}.  It is clear that as the mass of the galaxy increases, so does the efficiency of cold disk formation.  For the DC14 model, the evolution is rather weak and $f_d\propto {\rm M_{vir}^{0.16}}$.  There is still a reasonably large amount of scatter around this relation but the general trend persists.  For the NFW model, the RANSAC algorithm finds a fit where the efficiency of disk formation also increases with galaxy mass.  However there is so much scatter in this relation that the RANSAC algorithm classifies many of the points as outliers and this fit is not particularly trustworthy as there is no clear underlying relation.  Within the uncertainties, the slopes of the DC14 and NFW relations are consistent.  This is not particularly surprising since in certain regimes of M$_*$/M$_{\rm halo}$, the DC14 model reduces to NFW.  Outside of this regime, the DC14 model provides statistically better fits to the rotation curves than the NFW model \citep{Katz2016}: these galaxies are therefore outliers in the relation for the NFW model, driving our fitted relations closer to those of the DC14 model. These results agree with the $M_\text{b}-M_\text{vir}$ relation derived by combining the stellar mass abundance matching of \citet{Moster2013} with the $M_*-M_\text{gas}$ relation of \citet{Dutton_2011}.

\begin{table}
\centering
\begin{tabular}{@{}lccccc@{}}
Halo Model & $A$ & $\sigma_A$ & $B$ & $\sigma_B$ & $\sigma_{f_d}$ \\
 \hline
DC14 & 0.16 & 0.09 & -2.55 & 1.02 & 0.14 dex \\
NFW  & 0.26 & 0.09 & -3.74 & 1.02 & 0.24 dex\\
\hline
\end{tabular}
\caption{Fitting parameters for $f_d$ as a function of ${\rm M_{vir}}$ as given in Equation~\ref{fdmvir} along with their uncertainties.  $\sigma_A$ and $\sigma_B$ are the $1-\sigma$ uncertainties on the slope and normalisation of relation calculated from the 10,000 RANSAC fits to the resampled catalogues.  $\sigma_{f_d}$ is the scatter in $f_d$ calculated as 1.48 times the average median absolute deviation of the inliers of the 10,000 RANSAC fits.}
\label{tabfd}
\end{table}

\begin{figure*}
\centerline{\includegraphics[scale=1]{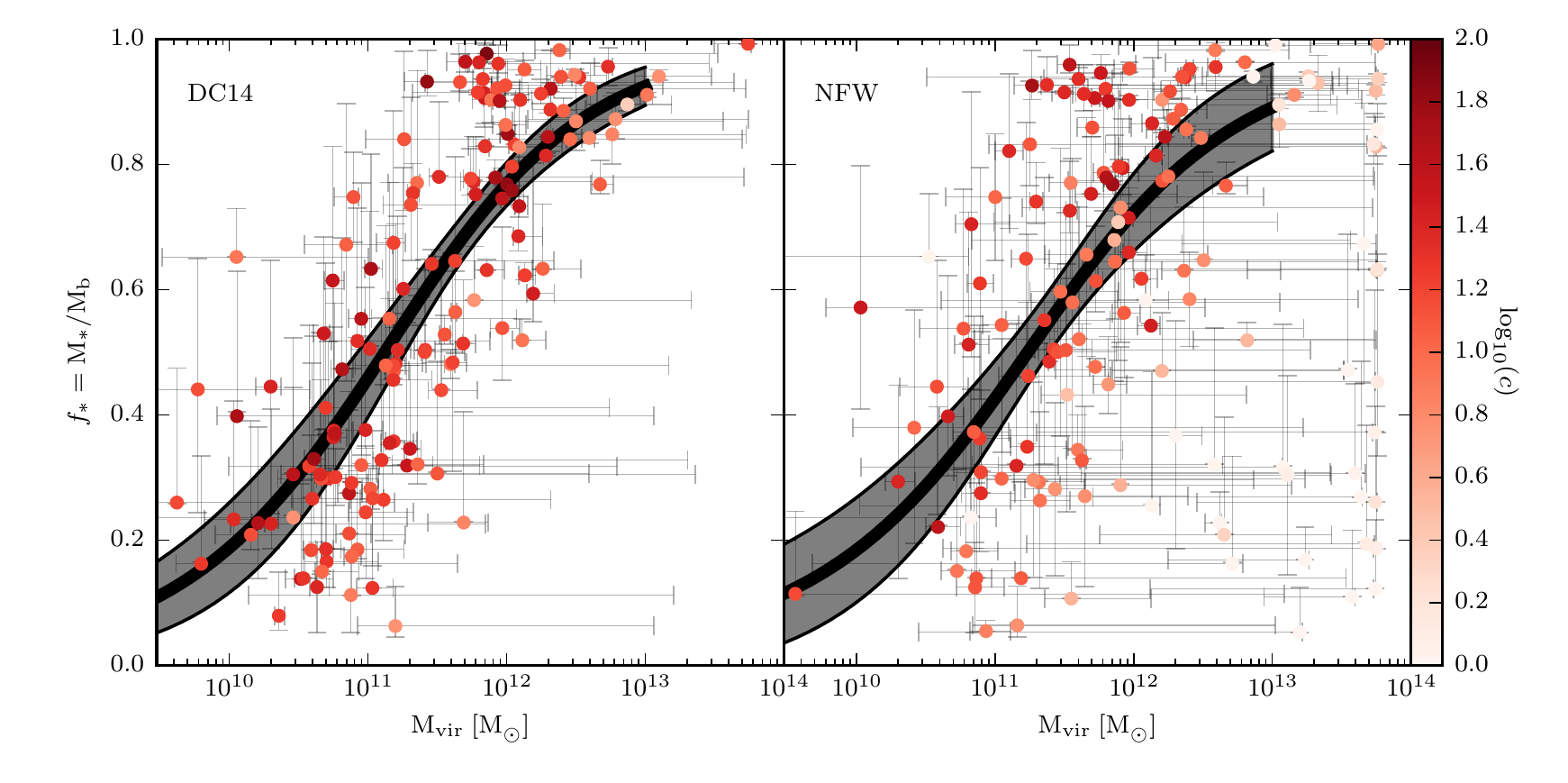}}
\caption{$f_*$ as a function of ${\rm M_{vir}}$ for the DC14 model (left) and the NFW model (right). The points represent the maximum likelihood fits to the observed rotation curves and they are coloured by the log of the concentration.  The error bars represent the $2\sigma$ uncertainties on the parameters for individual galaxies.  The thick black line shows the best fit values to the function given in Equation~\ref{logit} as determined by the RANSAC algorithm. The grey shaded region shows the $1\sigma$ uncertainty on this fit.}
\label{fs}
\end{figure*}

Interestingly, the trend in $f_d$ never comes close to unity, indicating that most of the baryons in late-type galaxies do not reside in cold disks, consistent with many other works in the literature \citep{Fukugita1998,Danforth2005,Shull2012}. Rather, $f_d\ll1$ implies that galaxy formation is a globally inefficient process. 

Possible scenarios that may lead to $f_d<1$ include: 1) the baryons never cooled and settled into a disk in the first place, or 2) the baryons settled into a disk and then feedback processes heated them up or ejected them from the galaxy.  Of course, these two scenarios are not mutually exclusive. There is evidence for galactic winds in many star forming galaxies \citep[e.g.][]{Veilleux2005} and the presence of metals in the CGM confirms a degree of recycling and ejection of gas from inside the galaxy \citep[e.g.][]{Tumlinson2017}.  Likewise the observation of hot halo gas surrounding galaxies indicates that there is a substantial heating process \citep[e.g.][]{Anderson2011}, although this may be due to virial shocks. It is important to note however that significant controversy still remains about the ubiquity of winds in star-forming systems, with some studies finding little evidence for them~\citep{Lelli_starburst, Concas}. If the mechanism that we describe here is responsible for setting the cold baryon fractions of late-type galaxies, we would expect some observational signatures of gas motion, at least in the most massive systems.

For the lowest mass galaxies, ${\rm M_{vir}\sim10^9\ M_{\odot}}$, the virial temperatures approach the temperature of ionized gas ($\sim10,000$~K) and reionization may have prevented accretion of gas onto some of these galaxies \citep{Gnedin2000,Okamoto2008}.  For more massive galaxies, this should not be an issue, although stellar and AGN feedback can still inhibit accretion onto a galaxy \citep[e.g.][]{Mitchell2017}.  In the remainder of this work we will assume the ``worst-case'' scenario where all baryons that could have accreted onto the galaxy have done so, and subsequent feedback processes are entirely responsible for heating the ``missing baryons'' into a hot halo or ejecting them from the galaxy. There are many effects which may prevent gas accreting onto a halo; we discuss this assumption further and the effect it has on the required feedback strength in Section~\ref{caveats}.

In Figure~\ref{fs}, we plot $f_*$ as a function of ${\rm M_{vir}}$ for both the DC14 and NFW halo models. DC14 haloes convert as low as $10\%$ and up to $60\%$ of their available baryons into stars at ${\rm M_{vir}}\sim10^{11}\ {\rm M_{\odot}}$, and $\sim50-100\%$ at ${\rm M_{vir}}\sim10^{12}\ {\rm M_{\odot}}$. Note that the scatter about the relation is very large even though the general trend is an increasing $f_*$ with virial mass. The galaxies with the lowest relative baryon content are also the most inefficient at forming stars. Once again, we see a similar trend for the NFW model, although the scatter is larger.

We fit the galaxy formation efficiency as a function of ${\rm M_{vir}}$ to a logistic function (sigmoid curve) of the form:
\begin{equation}
    f_*=\frac{1}{1+e^{-C\log_{10}({\rm M_{vir}}/M_0)}}.
    \label{logit}
\end{equation}
We constrain both $C$, which controls the steepness of the interpolation, and $M_0$, which is the point where the galaxy formation efficiency is $50\%$. The procedure for this fit is exactly analogous to what was done for the $f_d-{\rm M_{vir}}$ relation: we create 10,000 mock catalogs with both $f_*$ and ${\rm M_{vir}}$ computed from the MCMC chains, and use the RANSAC algorithm to find the best-fitting logistic function. The fitting parameters reported in Table~\ref{sefftab} are calculated as the means over the 10,000 fits, and the uncertainties are the $1\sigma$ deviations. In Figure~\ref{fs} we show the derived relation and confidence intervals as the thick black line and grey shaded region respectively.

\begin{table}
\centering
\begin{tabular}{@{}lccccc@{}}
Halo Model & $C$ & $\sigma_C$ & $\log_{10}(M_0)$ & $\sigma_{\log_{10}(M_0)}$ & $\sigma_{f_*}$ \\
 \hline
DC14 & 1.38 & 0.296 & 11.10 & 0.17 & 0.38 \\
NFW & 1.28 & 0.388 & 11.21 & 1.13 & 0.46 \\
 \hline
\end{tabular}
\caption{Fitting parameters for $f_*$ as a function of ${\rm M_{vir}}$ as given by Equation~\ref{logit}. $\sigma_{f_*}$ is the scatter in $f_*$ calculated as 1.48 times the average median absolute deviation of the 10,000 fits}
\label{sefftab}
\end{table}
We find that $M_0$ and $C$ are similar between the two different halo models, and that galaxies with ${\rm M_{vir}}\sim10^{11.1}$~M$_{\odot}$ convert $\sim50\%$ of available baryons into stars.

\section{Theoretical Framework}
\label{theory}
SN feedback can in principle impart both momentum and heat to the gas \citep{Taylor1950,Sedov1959}, each of which can play a role in regulating galaxy growth.  If the mechanical feedback is strong enough that gas clouds reach the escape velocity of the halo (${\rm V_{esc}}$), then, absent significant hydrodynamical drag from the ISM, the gas may be completely ejected from the halo. Unless the gas is re-accreted, it is then unable to undergo future star formation. Alternatively, instead of ejecting gas from the halo, SN feedback may simply heat it to near the virial temperature of the halo, where it can either bubble off of the galaxy or be recycled onto the disk~\citep{Brook2014,Christensen2016}. Both of these mechanisms likely play a part in regulating star formation.

We calculate the energy needed to eject the ``missing baryons" from the galaxy ($E_{\rm ej}$) or heat them to the virial temperature  ($E_{\rm heat}$) as a function of the halo mass of the galaxy (${\rm M_{halo}}$), as follows:
\begin{equation}
    E_{\rm ej}=\frac{1}{2}{\rm M_{X}}{\rm V_{esc}}^2.
    \label{Eejeqn}
\end{equation}
and
\begin{equation}
    E_{\rm heat}=\frac{3}{2}{\rm M_X}\frac{k_B({\rm T_{vir}-T_{ini}})}{\mu m_p}.
    \label{Eheat}
\end{equation}
where ${\rm M_X}$ is the gas mass which we wish to heat or eject, ${\rm T_{vir}}$ is the virial temperature of the halo, ${\rm T_{ini}=10,000~K\ (100\ K)}$ is the initial temperature of ionized (neutral) gas, $k_b$ is the Boltzmann constant, $m_p$ is the proton mass, and $\mu=0.59\ (1.22)$ is the mean molecular weight of the ionized (neutral) gas.

These equations are necessarily very simplified descriptions of gas ejection and heating: Equation~\ref{Eejeqn} assumes that the gas cloud resides initially near the centre of the halo, that the halo is in the steady state and not formed hierarchically, that there are no energy losses, and that there is no hydrodynamical drag. While the former two effects would cause overestimation of the energy required for the gas to leave the halo, the latter two would cause the converse. We therefore retain this form for simplicity. Similarly, Equation~\ref{Eheat} is subject to vagaries in heat transport within the gas between the patch heated specifically by the SNe and the rest of the ISM. Our oversimplification may be most evident in the case of a cold metal-rich cloud moving at high speed through a hot corona: in this case the differences in metallicities and densities may cause cause cooling of the coronal gas by the Kelvin-Helmholtz instability~\citep{Marinacci} Nevertheless, we believe Equations~\ref{Eejeqn} and~\ref{Eheat} will give a reliable intuition into the energy scales required to regulate a galaxy.

We aim to apply this model to both the NFW and DC14 models, which are subsets of the more general ($\alpha,\beta,\gamma$) profile.  For this reason, we have derived a series of analytic expressions that allow for quick evaluation of various quantities for the ($\alpha,\beta,\gamma$) density profile without requiring numerical integrals\footnote{See \cite{Dekel2017} for an alternative model which also avoids such integrals and provides reasonable fits to rotation curves.}.  These can be found in Appendix~\ref{app1}.

\subsection{Heating or ejecting the baryons}
We now turn to ${\rm T_{vir}}$, given by
\begin{equation}
    {\rm T_{vir}}=(\mu m_p/2k_B){\rm V_{vir}}^2.
\end{equation}
If ${\rm T_{vir}\gg T_{ini}}$, which is true for galaxies with ${\rm M_{vir}\gtrsim2\times10^9\ M_{\odot}}$ if the gas is ionized and always true if the gas is neutral, then $E_{\rm heat}\propto {\rm M_XM_{vir}^{2/3}}$.  If we plug in our scaling for ${\rm V_{esc}}$, we also find that $E_{\rm ej}\propto {\rm M_XM_{vir}^{2/3}}$.  Interestingly, the scaling is the same for both quantities.  We can compare this to the energy available from SN feedback ($E_{\rm fb}$) which is given by
\begin{equation}
    E_{\rm fb}({\rm M_{vir}})=E_{\rm SN}\epsilon_cf_{\rm SN}f_d({\rm M_{vir}})f_*({\rm M_{vir}})\frac{\Omega_{\rm b}}{\Omega_{\rm m}}{\rm M_{vir}},
    \label{eqnEfb}
\end{equation}
where $E_{\rm SN}=10^{51}$~ergs is the energy of an individual SN explosion, $\epsilon_c$ is the fraction of that energy that couples to the ISM and contributes to driving a wind, $f_{\rm SN}\sim0.01$~M$_{\odot}^{-1}$ is the number of SN explosions per unit solar mass \citep{Salpeter1955} and is dependent on the stellar IMF, $f_d({\rm M_{vir}})$ is the fraction of the mass of the galaxy with respect to the cosmological value that settles into a cold disk
\begin{equation}
f_d={\rm (M_{gas,cold}+M_*)/([\Omega_b/\Omega_{m}]M_{vir})},
\end{equation}
and $f_*({\rm M_{halo}})$ is the fraction of baryons that are part of the disk and form stars
\begin{equation}
f_*={\rm M_*/(M_{gas,cold}+M_*)}.
\end{equation}
Thus
\begin{equation}
f_df_*\frac{\Omega_{\rm b}}{\Omega_{\rm m}}{\rm M_{vir}=M_*}.
\end{equation}

A priori, both $f_d$ and $f_*$ are unknown and both $E_{\rm SN}$ and $f_{\rm SN}$ are expected to be constant\footnote{This assumes that the stellar IMF is constant as a function of halo mass.}. For simplicity, we assume that the efficiency with which SN couple to the ISM ($\epsilon_c$) is independent of galaxy mass. For SN feedback to be the dominant mechanism which regulates galaxies in the mass range $10^{12}\gtrsim {\rm M_{vir}}\gtrsim10^9\ {\rm M_{\odot}}$, one must have ${\rm M_*}\propto {\rm M_XM_{vir}^{2/3}}$.

For our back-of-the-envelope calculation, the quantity ${\rm M_X}$ is well approximated as ${\rm M_X\sim M_b\sim (\Omega_b/\Omega_m)M_{vir}}$, where ${\rm M_b=M_*+M_{gas,cold}+M_{gas,hot}}$ is the baryonic mass of the galaxy.  Substituting in for ${\rm M_X}$ and ${\rm M_{vir}}$ in our scaling relation, we find
\begin{equation}
    {\rm M_*\propto M_{vir}^{5/3}},
\end{equation}
assuming that all of the baryons are either heated to the virial temperature of the galaxy or completely ejected, solely due to SN feedback.  What we have just derived is the theoretical slope of the ${\rm M_*/M_{halo}}$ relation assuming that SN feedback is the dominant mechanism which regulates galaxies, and that the majority of baryons are not in stars or cold gas. If higher mass galaxies have fewer missing baryons compared to low mass galaxies the slope will be less than $5/3$, and vice versa.

\begin{figure*}
\centerline{\includegraphics[trim={0 1.8cm 0 0},clip,scale=1]{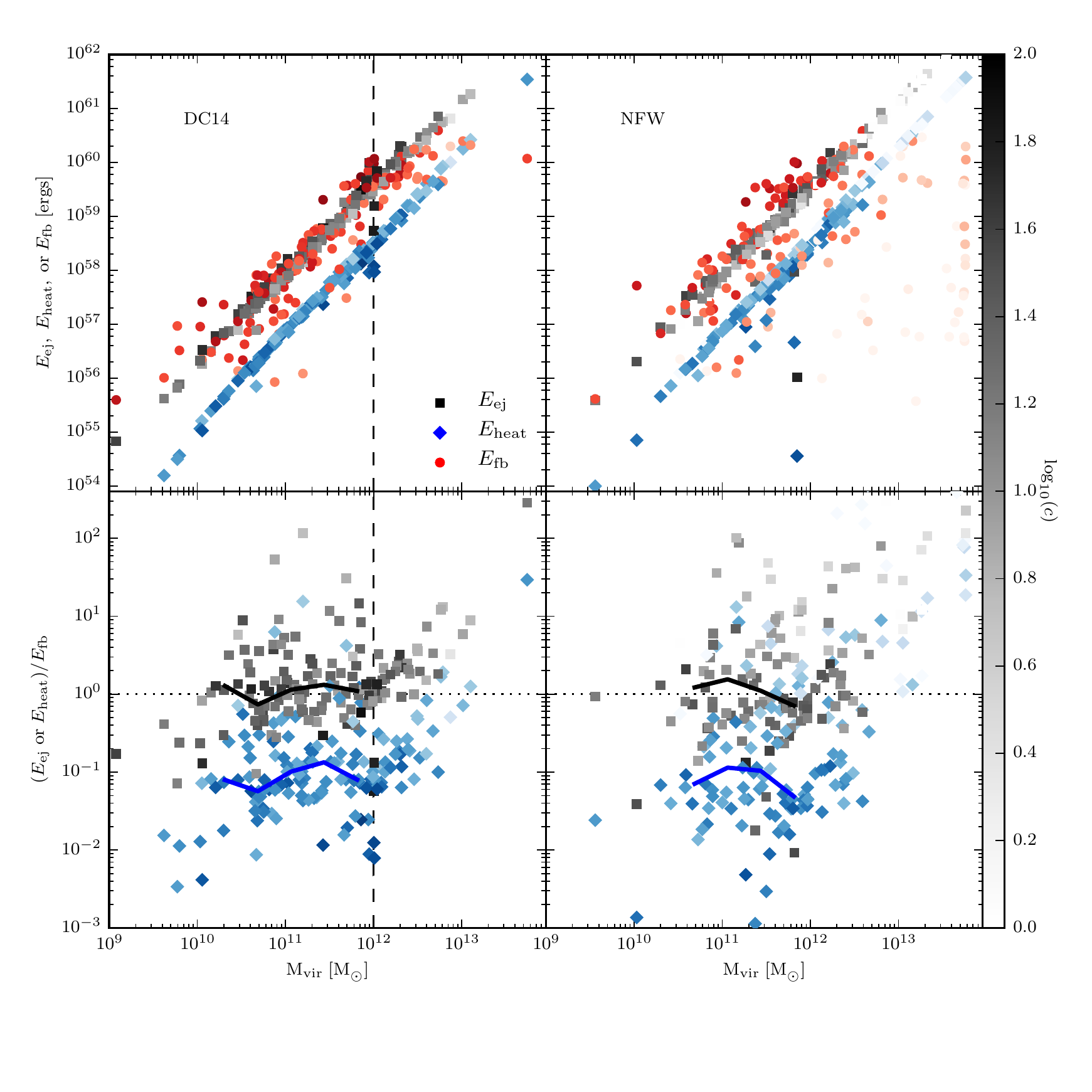}}
\caption{The top panel shows a comparison of energy available from SN, when all the energy is coupled to the ISM, i.e. $\epsilon_c=1$ (red points), with the energy required to heat the ``missing baryons" (blue points), or eject the ``missing" baryons from the galaxy (black points) as a function of ${\rm M_{vir}}$.  The shading of the points is by the log of the concentration. The vertical dashed line represents the mass at which the DC14 model is expected to break down. The bottom panel shows the ratio of $E_\text{ej}$ or $E_\text{heat}$ to $E_{\rm fb}$; a value of 1 (shown the horizontal dotted line) indicates that SN energy is exactly sufficient to remove the gas or heat it to ${\rm T_{vir}}$.  The black and blue lines shows the median fraction of the SN energy that needs to couple to the gas (in bins of $\Delta\log_{10}({\rm M_{vir}})=0.375$ for systems with ${\rm M_{vir}<10^{12}M_{\odot}}$) in order to eject or heat the baryons to the virial temperature of the halo, respectively.  Bins with fewer than five galaxies are not shown.}
\label{energycomp_ind}
\end{figure*}

\section{Comparison with Observation}
\label{comparison}
\subsection{Individual Galaxies}
With both $f_d$ and $f_*$ measured for 147 galaxies from the SPARC data set, along with their masses and concentrations derived in \cite{Katz2016}, we can now compare $E_{\rm ej}$, $E_{\rm heat}$, and $E_{\rm fb}$.  In the top row of Figure~\ref{energycomp_ind}, we plot the energy required to eject or heat the ``missing baryons" for each individual galaxy in our sample and compare it to the energy available in SN feedback. Here, we show the case where $\epsilon_c=1$ (i.e. that all SN energy contributes to ejecting or heating the gas). We caution however that this is the most optimistic case. Not only will some SN energy fail to be transformed into kinetic energy, but gas along the plane of the disk will be less likely to escape if the outflow is primarily perpendicular. Hydrodynamical models predict that this may result in $\epsilon_c$ values significantly less than unity~\citep{Rogers2013,Walch2015,Gentry2017}. In addition, we have assumed that all the supernova energy couples to the central regions of a halo to make a core. If this is not true the efficiency of the feedback for core formation will be reduced by a further factor. Since $E_{\rm fb}$ depends linearly on $\epsilon_c$, it is trivial to transform our results to other values. In the bottom row of Figure~\ref{energycomp_ind} we show the ratios of $E_{\rm fb}/E_{\rm heat}$ and $E_{\rm fb}/E_{\rm ej}$ which gives insight into the exact values of $\epsilon_c$ needed to either eject or heat the missing baryons.

For the DC14 model, the scatter is relatively tight.  There are a few outliers as expected, but many of these have large uncertainties on the halo fits \citep{Katz2016}. We expect variation for individual galaxies because galaxy properties such as disk size, metallicity, star formation history, merger history, and environment may cause deviations from our simple model.  It is however promising that we find relatively good agreement between the scalings of $E_\text{fb}$ and $E_\text{heat}$ or $E_\text{ej}$ and that the scatter for individual galaxies is not huge.

There is significantly more scatter for the NFW model than for DC14. In Figure~\ref{energycomp_ind}, the points which represent $E_{\rm fb}$ are merely a rescaling of the ${\rm M_*/M_{halo}}$ relation presented in \cite{Katz2016}, where it was shown that the scatter in this plane was too large for the NFW model to be consistent with observations.  For the NFW model, the scatter in $E_{\rm ej}$ and $E_{\rm heat}$ are dependent on the scatter in $c$ through ${\rm V_{esc}}$ and weakly dependent on the scatter in ${\rm M_b}$ since for the most part ${\rm M_{vir}\gg M_b}$.  The scatter in $c$ is clearly smaller than the scatter in ${\rm M_*}$ which is why there is much more observed scatter in  $E_{\rm fb}$ for the NFW model compared to the scatter in $E_{\rm ej}$ and $E_{\rm heat}$.  This is not true for the DC14 model where the scatter in ${\rm M_*}$ is much smaller.

\begin{figure*}
\centerline{\includegraphics[scale=1]{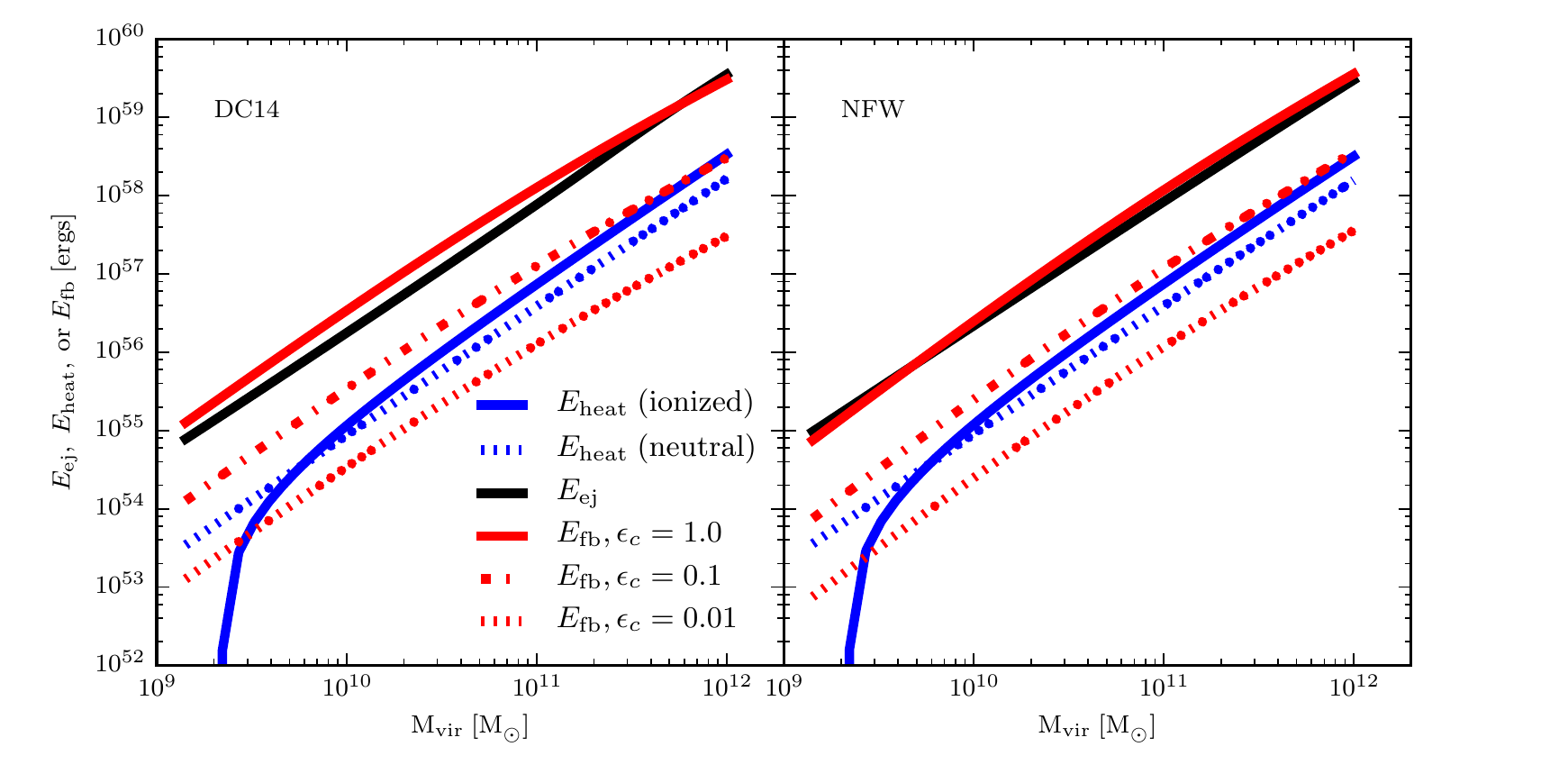}}
\caption{Comparison of energy available from SN feedback (red lines) with the energy required to heat the ``missing" baryons (blue), or eject the ``missing" baryons from the galaxy (black) as a function of ${\rm M_{vir}}$.  The different line styles for the red lines represent different coupling efficiencies of the SN energy to the ISM.  The different line styles for the blue lines represent if the gas starts out ionized or neutral.}
\label{energycomp}
\end{figure*}

Looking at galaxies with ${\rm M_{vir}>10^{12}\ M_{\odot}}$ which corresponds to stellar mass just under $10^{11}~{\rm M_{\odot}}$, there is very tentative evidence for a slight turnover in the energy available from SN feedback while the energy needed to eject or heat the ``missing baryons" continues to increase.  At these masses, we are using an extrapolation of the DC14 model so our interpretation needs to be taken with caution. This would however be consistent with the idea that at these masses other feedback mechanisms are starting to play a role \citep{Schaye2015}.  This picture is in line with the observed luminosity function where we see a turnover in the slope at these masses \citep{Li2009, Bernardi_SMF}.

Since the red points overlap the black points in the left panel of Figure~\ref{energycomp_ind}, 100\% of the energy available from SN would need to couple to the gas in order to eject all of the ``missing baryons" from the halo.  This value is rather unrealistic as it is well established that this coupling is inefficient \citep{Walch2015,Rogers2013} and dependent on the density of the ambient medium \citep{Cowie1981}.  As has been shown before (e.g.~\citealt{Yepes,Efstathiou}), more energy is required to eject the ``missing baryons'' from the galaxy than to heat them to ${\rm T_{vir}}$.  Less than $\sim10\%$ of the SN feedback energy needs to couple to the ISM in order to heat the gas to the virial temperature and have the red points overlap the blue points in the left panel of Figure~\ref{energycomp_ind}.  This value is strongly degenerate with the chosen stellar IMF: for more top heavy IMFs such as Chabrier \citep{Chabrier2003} the required coupling would be less than the value derived here.  Nevertheless, it is likely that more of the ``missing baryons'' are located in a hot halo surrounding the galaxy than that they were ejected, even for our lowest mass systems of $\sim10^9\ {\rm M_{\odot}}$.  The key aspect of this diagram is that the slope of the red points is consistent with the slope of the black and blue points indicating that the amount of feedback energy available is a simple rescaling of the amount needed to heat or eject the ``missing baryons".  This can be seen in the bottom panels of Figure~\ref{energycomp_ind} where we plot the median coupling efficiency, in bins of halo mass between $10^{10}<{\rm M_{vir}/M_{\odot}}<10^{12}$, required to heat or eject the ``missing baryons".  This value is roughly constant as a function of halo mass.  At lower masses, there are few galaxies in our sample and thus the median may not be truly representative. If SN are the dominant feedback mechanism regulating the galaxies, our work suggests that the coupling efficiency is constant across a wide range of halo masses.

In summary, even with 100\% coupling efficiency of SN energy, there is often not enough energy to eject all of the ``missing baryons" from the galaxy.  For the DC14 model, only $\sim9\%$ of the total energy available from SN needs to couple to the gas to heat it to the virial temperature of the halo. We find that this value is reasonably independent of halo mass. We remind the reader however that our model is idealised: we have assumed that galaxies initially accrete the cosmological baryon fraction, all supernova energy couples with constant efficiency to dark matter in the central regions, and that halos begin with an NFW profile. More detailed conclusions will require hydrodynamical simulations where these assumptions are not necessary.

\begin{figure*}
\centerline{\includegraphics[scale=1]{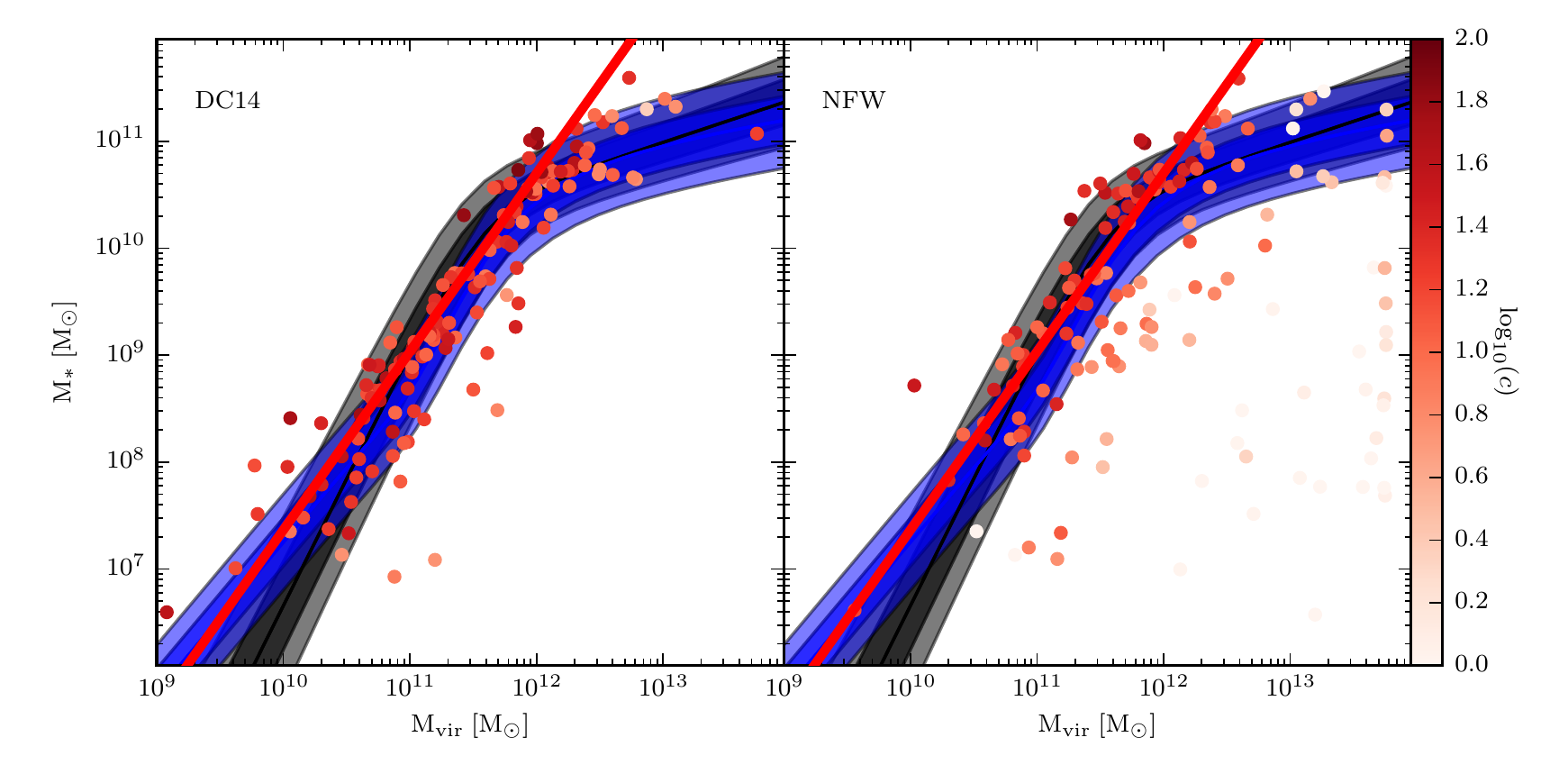}}
\caption{${\rm M_*/M_{halo}}$ relation of the maximum posterior fits for each individual galaxy in our sample compared to the predicted relation from \protect\cite{Moster2013} (grey shaded region) and \protect\cite{Behroozi2013} (blue shaded region).  The shaded regions map out the $1\sigma$ and $2\sigma$ scatter in the relations.  Points are coloured by $\log_{10}(c)$.  The red line shows an arbitrary scaling of ${\rm M_*\propto M_{vir}^{5/3}}$.  For ${\rm M_{vir}}>10^{12}\ {\rm M_{\odot}}$, the $5/3$ scaling not longer well represents the data which may indicate that other processes, such as AGN, might be affecting these galaxies.}
\label{mostbeh}
\end{figure*}

\subsection{Mean Relations}
Rather than perform this exercise for individual galaxies, we can use the mean relations for $f_d$ and $f_*$ derived in Section~\ref{obscon}. In Figure~\ref{energycomp}, we plot the energy needed to either heat the ``missing baryons" or eject them completely from the halo for both the DC14 and NFW models, and compare this to the energy available in SN feedback for three different constant values of $\epsilon_c$.  The mass in ``missing baryons" is simply
\begin{equation}
    {\rm M_X}=(1-f_d)\frac{\Omega_{\rm b}}{\Omega_{\rm m}}{\rm M_{\rm vir}}.
\end{equation}
To calculate $E_{\rm ej}$, we assume that all galaxies lie on the ${\rm M_{halo}}-c$ relation as given in \cite{Dutton2014}.  Furthermore, to calculate ${\rm V_{esc}}$ for the DC14 model, we need to know $\alpha,\ \beta,\ {\rm and}\ \gamma$ which can make up to $\sim7\%$ difference in this velocity (see Figure~\ref{vesccomp}). These can be determined from our fits of $f_d$ and $f_*$, given the scaling relations presented in DC14. We plot two different quantities for $E_{\rm heat}$, one assuming that the gas which is being heated starts out ionized (${\rm T_{ini}=10,000\ K}$ and $\mu=0.59$), and the other that the gas begins neutral (${\rm T_{ini}=100\ K}$ and $\mu=1.22$).

Figure~\ref{energycomp} shows that $E_{\rm fb}$, $E_{\rm ej}$ and $E_{\rm heat}$ scale in the same way with ${\rm M_{vir}}$ for both the DC14 and NFW models. The relation between $f_d$ and ${\rm M_{vir}}$ has a large scatter and for the NFW model, it is not clear that the function we have fit to the data is truly representative.  This likely stems from the fact that the NFW halo profile does not provide good fits to the rotation curves of SPARC galaxies \citep{Katz2016}, and therefore it is unwise to draw strong conclusions about the NFW halo from these mean relations.  For the DC14 model, there is a much clearer indication that $f_d$ increases with ${\rm M_{vir}}$, and that the slope is roughly linear in log-space.  It is encouraging that the DC14 model naturally predicts the scaling one would expect if SN feedback is the dominant mechanism regulating star formation and gas cooling in galaxies.

\subsection{Comparison to other ${\rm M_*/M_{halo}}$ relations}
In \cite{Katz2016}, it was claimed that the maximum-likelihood parameters for the DC14 halo model fits adhered well the the ${\rm M_*/M_{halo}}$ relation predicted by \cite{Moster2013}, even without $\Lambda$CDM priors. For the mass range we consider here (${\rm 10^{12}\gtrsim M_{vir}/M_{\odot}\gtrsim10^9}$), \cite{Moster2013} find ${\rm M_*\propto M_{halo}^{2.376}}$. This scaling is clearly different that the relation of ${\rm M_*\propto M_{halo}^{5/3}}$ predicted for galaxy regulation by SN. How then can our results be consistent with both?

In Figure~\ref{mostbeh}, we plot the ${\rm M_*/M_{halo}}$ relation of the maximum-likelihood fits from each galaxy in our sample compared to the predicted relations from \cite{Moster2013} and \cite{Behroozi2013}.  As was shown in \cite{Katz2016} the fits for the DC14 model adhere reasonably well to the $2\sigma$ scatter in the relation from \cite{Moster2013}.  Overplotted on this diagram is a red line with a scaling of ${\rm M_*\propto M_{vir}^{5/3}}$ and an arbitrary normalisation.  Note that this is not a fit to the data.  It is clear that the ${\rm M_*/M_{halo}}$ predicted by \cite{Moster2013} is slightly steeper than this relation, although within $2\sigma$ they are not in terrible disagreement. Although we find a different slope, we maintain approximate agreement with both relations.

\begin{figure*}
\centerline{\includegraphics[scale=1]{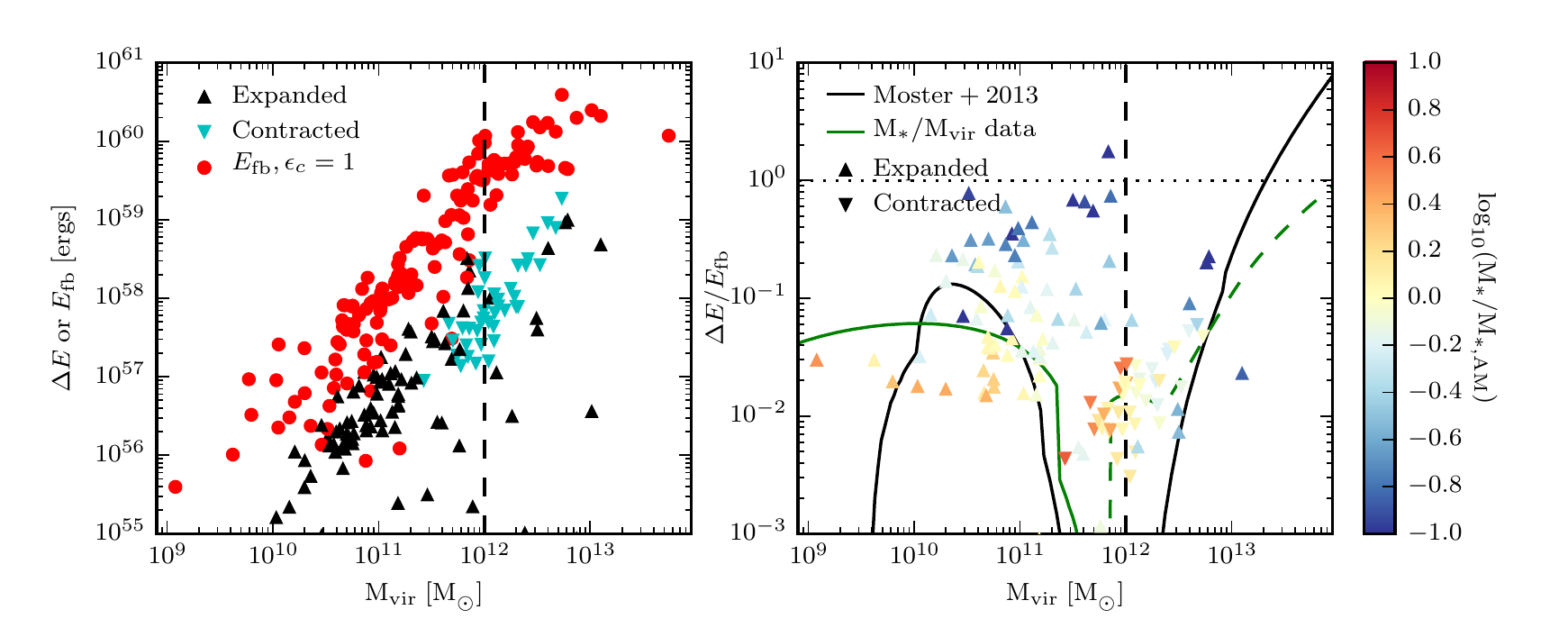}}
\caption{(Left) Energy needed to transform the halo from NFW to DC14 (black and cyan) versus the energy available in feedback (red).  Black points represent that this process requires energy (i.e. that the DC14 model has removed mass from the centre compared to the primordial NFW halo) while cyan points represent contracted haloes and thus the energy difference is negative (i.e. mass has moved closer to the halo centre). The available feedback energy is shown for $\epsilon_c=1$.  (Right) The ratio of the energy needed to restructure the halo to the maximum energy available from SN. Upwards and downwards triangles represent cored and adiabatically contracted haloes, respectively. Galaxies falling below the dotted line have enough SN energy to either form a core or counter-balance adiabatic contraction. Points are color-coded by the logarithmic distance from the empirically determined ${\rm M_* - M_{vir}}$ relation. Red points have higher than average stellar mass while blue points have lower than average stellar mass. The solid black line shows the expectation for galaxies that fall on the ${\rm M_* - M_{vir}}$ relation and the ${\rm M_{vir}} - c_{\rm vir}$ relation from \protect\cite{Moster2013} and \protect\cite{Maccio2008}, respectively.  The green line is the same as the black line but considers the empirically determined ${\rm M_* - M_{vir}}$ and ${\rm M_{vir}} - c_{\rm vir}$ relations from halo fits to the SPARC rotation curves. The solid and dashed portions of these curves indicates halos that have been expanded and contracted, respectively. Galaxies that follow the ${\rm M_* - M_{vir}}$  relation from \protect\citet{Moster2013} and the ${\rm M_{vir}} - c_{\rm vir}$ relation from \protect\cite{Maccio2008} should always exhibit some degree of central expansion.}
\label{Erest}
\end{figure*}

The ${\rm M_*/M_{halo}}$ relation of \cite{Behroozi2013} is shallower at the low-mass end than \cite{Moster2013} and is therefore in better agreement with our predicted slope. Our normalisations tend to be slightly higher than \cite{Behroozi2013}, although the disagreement is marginal at best.  We should note however that their stellar masses are derived in a different band than the SPARC galaxies.  There is a much larger uncertainty in ${\rm M_*/L}$ in visual bands than at $3.6\mu$ \citep{Mcgaugh2014} and there can be systematic differences between bands. Any disagreement between the normalisation of the DC14 model and these relations is marginal and can be readily reconciled with a change in ${\rm M_*/L}$.  Finally, the relations derived by \cite{Behroozi2013} and \cite{Moster2013} consider all galaxies, whilst we consider only late-type galaxies. Part of the difference in slope may well derive from morphological selection effects.

If we measure the slope of the $\log_{10}({\rm M_*})-\log_{10}({\rm M_{vir}})$ relation directly from the data for systems with ${\rm M_{vir}<10^{12}M_{\odot}}$, we find this value to be $1.51\pm0.13$ for the DC14 model and $1.67\pm0.20$ for the NFW model. As expected, the slope is slightly less than $5/3$ for DC14 because there is a weak scaling of $f_d$ with galaxy mass.  For the NFW case the slope is $5/3$, albeit with a significant number of outliers.  Note that ``preventative feedback" which is biased towards low mass galaxies (e.g. radiative feedback from reionization) would lead to an $f_d$ scaling that would also cause the slope of this relation to become shallower than $5/3$.

\section{Discussion}
\label{discussion}
\subsection{The effect of cooling}
It is clear from the previous section that there is not enough energy to eject all of the ``missing baryons" out of the galaxy. If the gas is only heated to ${\rm T_{vir}}$ and remains in the halo, it has a chance to cool and settle back onto the disk.  The simulations of \cite{Christensen2016} find that $\sim50\%$ of material kicked out of the cold disk due to SN feedback is re-accreted on a timescale of $\sim1$~Gyr, independent of halo mass.  This would imply that in order to maintain the gas at the virial temperature of the halo, one would have to couple $50\%$ more energy than we concluded from Equation~\ref{Eheat}.

For our individual galaxies, a coupling efficiency of $\sim9\%$ is roughly what is required to heat all of the ``missing baryons" to ${\rm T_{vir}}$.  To maintain this temperature if cooling is included, one would then require $\sim13.5\%$ (assuming 50\% of the baryons are re-accreted) of the total available energy from SN to couple thermally to the ISM.  Once again, this is degenerate with the choice of stellar IMF.  Assuming a Chabrier IMF \citep{Chabrier2005} would decrease this value by a further $\sim20\%$.  This of course neglects any energy input pre-SN which may come in the form of stellar winds, radiation pressure, or photoheating of species other than hydrogen.  Likewise, cosmic rays created in SN may play a role in driving an outflow \citep[e.g.][]{Breit1991}.  Furthermore, our estimates here are the absolute upper limits of the amount of energy one needs to account for the ``missing baryons" as they assume that all baryons which should be in the halo have, at one point, accreted onto the cold disk.  As has been shown from cosmological hydrodynamics simulations, the feedback energy can eject baryons as well as prevent accretion onto the halo which would reduce the amount of energy needed \citep[e.g.][]{Mitchell2017}.  For the more massive haloes, we may expect strong virial shocks.  The actual amount of baryons that actually cool onto the disk is likely to be closer to $50\%$, independent of halo mass \citep{Christensen2016}, bringing our absolute upper limit for the net $\epsilon_c$ back down to $\sim9\%$ if cooling is considered.  It is well known that radiative cooling is a very efficient processes, making the coupling of thermal energy relatively inefficient.  Our upper limit of $\epsilon_c\sim9\%$ puts our model in the realistic regime of what can be expected from simulations \citep[e.g.][]{Walch2015,Gentry2017}.

\subsection{Energy needed to restructure the halo}
Thus far, we have only been concerned with the energy required to remove the ``missing baryons" from the centres of their host haloes.  If we believe that the primordial haloes for these galaxies have NFW profiles, similar to \cite{Penarrubia2012}, we can calculate the change in gravitational potential energy needed to convert them to a DC14 profile:
\begin{equation}
    \Delta E = \frac{W_{\rm DC14}-W_{\rm NFW}}{2},
\end{equation}
where
\begin{equation}
    W = -4\pi G \int^{\rm R_{max}}_0\rho(r)M(r)rdr.
    \label{potdiff}
\end{equation}

In the low mass regime, DC14 reverts back to NFW so $\Delta E = 0$. At high masses, adiabatic contraction is expected to be the dominant effect, resulting in $\Delta E < 0$.  It is only at intermediate masses, between $10^6<{\rm M_*/M_{\odot}}<10^{10}$, where the inner slope is shallower than $-1$, that we expect mass to have been removed from the centre of the galaxy and hence $\Delta E > 0$.  \cite{Penarrubia2012} have set ${\rm R_{max}=R_{vir}}$ as the upper bound for the integral in Equation~\ref{potdiff}.  \cite{Maxwell2015} have pointed out that the upper bound for this integral should be set by a condition where both the mass and the density are equal between the two different halo density profiles.  There is no generic radius where this is true when comparing DC14 to NFW and thus we have chosen the radius where the mass profiles first converge.  Integrating out to ${\rm R_{vir}}$ does not change our results substantially and using our current method, we find that our results are consistent with both \cite{Maxwell2015,Brook2015}. 

In Figure~\ref{Erest}, we use the DC14 model fits to calculate $\Delta E$, and compare with the available energy from SN feedback assuming $\epsilon_c=1$.  Nearly all of the red points fall above the black points indicating that there is more energy available in the form of energetic feedback than is required to transform the halo.  This can more easily be seen in the right panel of Figure~\ref{Erest} where we show the ratio of SN energy to the amount required to restructure the halo.  At higher masses, adiabatic contraction takes over and $\Delta E$ becomes negative.  The points in the right panel have been colored by the ratio of their stellar mass to the empirically determined ${\rm M_* - M_{vir}}$ relation.  The points that scatter high tend to have stellar masses that are lower than average, while the red points tend to fall lower and have higher than average stellar masses.  

For the systems that exhibit cores, we find no mass strong dependence in the ratio of $\Delta E/E_{\rm fb}$.  The green curve is mostly flat between $10^9<{\rm M_{vir}}/M_{\odot}<3\times10^{11}$.  A median (IMF dependent) coupling efficiency of $\sim5\%$ is required for SN to be the dominant mechanism driving core formation.  There is once again significant scatter; however, this value is very consistent with the coupling efficiency that was obtained earlier when measuring how much energy was required to heat the ``missing baryons" to the virial temperature of the halo.  The  $\Delta E$ that we require to create cores is in good agreement with what is found in \cite{Maxwell2015} for their larger core models (see their Figure~3).  Note that our choice of ${\rm R_{max}}$ is often much larger than what is choses in \cite{Maxwell2015} and hence we can expect to require slightly more energy.  However, most of our systems do not exhibit central density slopes that are exactly zero and thus our energy requirement is lessened compared with the density profiles used in \cite{Maxwell2015} and thus we find a reasonable agreement.

Interestingly, the maximum energy available from SN is always greater than the energy loss during contraction. If the SN energy couples efficiently to the gas at a constant $\epsilon_c$, as suggested earlier, one may wonder how contraction is possible for such systems.  For this, we must understand how the dark matter responds to changes in the potential.  \cite{Pontzen2012} have shown that transformations of the dark matter halo require non-adiabatic changes in the potential, which depends on the star formation history. For instance, a galaxy that formed all of its stars in a single burst is likely to have a larger core than a galaxy with a similar stellar and halo mass but a more protracted star formation history.  It is well established that the specific star formation rate scales with stellar mass such that low mass galaxies have higher sSFRs compared to higher mass galaxies \citep[e.g.][]{Elbaz2007,Karim2011}.  Furthermore, simulations show that lower mass galaxies tend to have more bursty star formation histories compared to more massive galaxies \citep[e.g.][]{Brook2012}.  In the sample of galaxies presented here, a larger fraction of the higher mass galaxies exhibit contraction compared to lower mass systems which is consistent with this argument.  However, \cite{Penarrubia2012} have shown that the transformation is degenerate with formation redshift and other galaxy proprieties which may complicate our interpretation.  

In contrast to the cored galaxies, we find a mass dependence in the ratio of $\Delta E/E_{\rm fb}$ for the contracted systems such that $|\Delta E|/E_{\rm fb}\propto{\rm M_{vir}^{0.9}}$.  This indicates that more massive systems require a higher coupling efficiency in order to erase the contraction compared to lower mass systems.  The coupling efficiencies required to erase the contraction are rather low compared to what is required to create a core.  All contracted systems require a coupling efficiency $<10\%$ to erase the contraction.  Although there is always enough energy available from SN to complete this transformation, it remains to be determined how efficiently that energy eventually couples to the dark matter as a function of halo mass for these contracted systems.

\begin{figure*}
\centerline{\includegraphics[scale=1]{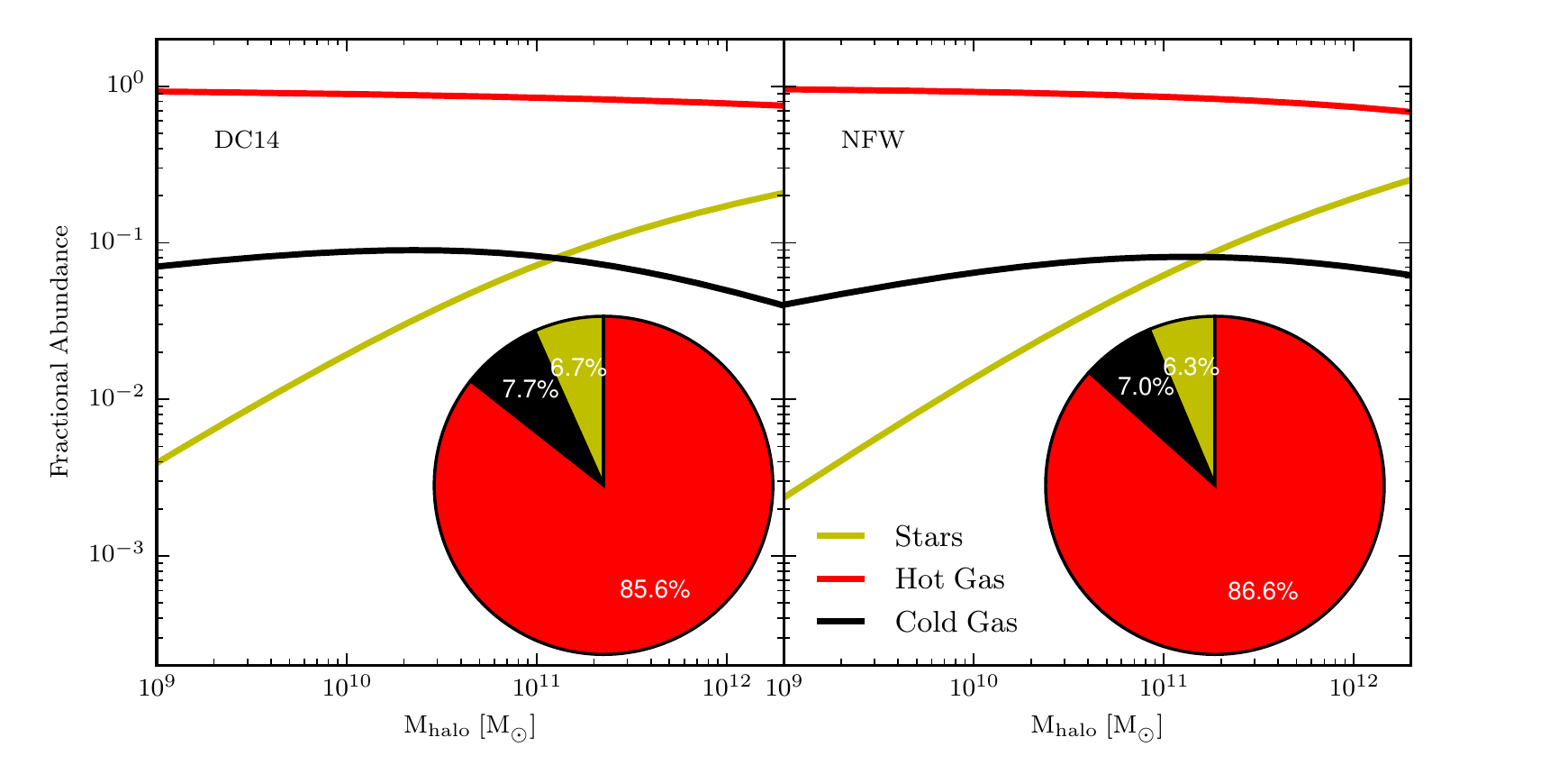}}
\caption{Fraction of cold gas (black), hot gas (red) and stars (yellow) as a function of ${\rm M_{halo}}$.  The inset pie charts show the fractional abundances integrated over the halo mass function.}
\label{barycen}
\end{figure*}

\subsection{Where are the baryons?}
\label{where}

Having formulae for $f_d$ and $f_*$ as functions of $M_\text{vir}$ allows us to take a census of the baryons in late type galaxies in various phases.  We have shown in Figure~\ref{energycomp} that it is unlikely that the ``missing baryons" have all been ejected from the haloes, but it is likely that many are in the hot phase surrounding galaxies.  The fractional abundances of hot gas (${\rm f_{M_{gas,hot}}}$), cold gas (${\rm f_{M_{gas,cold}}}$), and stars (${\rm f_{M_{*}}}$) in the mass range $10^9<{\rm M_{vir}/M_{\odot}}<10^{12}$ can be calculated as follows:
\begin{equation}
    {\rm f_{M_{gas,hot}}}=\frac{\int_{10^9\ {\rm M_{\odot}}}^{10^{12}\ {\rm M_{\odot}}}(1-f_d(M))M\frac{dn}{dM}(M)dM}{\int_{10^9\ {\rm M_{\odot}}}^{10^{12}\ {\rm M_{\odot}}}M\frac{dn}{dM}(M)dM},
\end{equation}
\begin{equation}
    {\rm f_{M_{gas,cold}}}=\frac{\int_{10^9\ {\rm M_{\odot}}}^{10^{12}\ {\rm M_{\odot}}}f_d(M)M\frac{dn}{dM}(M)dM}{\int_{10^9\ {\rm M_{\odot}}}^{10^{12}\ {\rm M_{\odot}}}M\frac{dn}{dM},(M)dM}-{\rm f_{M_*}}
\end{equation}
and
\begin{equation}
    {\rm f_{M_{*}}}=\frac{\int_{10^9\ {\rm M_{\odot}}}^{10^{12}\ {\rm M_{\odot}}}f_*(M)f_d(M)M\frac{dn}{dM}(M)dM}{\int_{10^9\ {\rm M_{\odot}}}^{10^{12}\ {\rm M_{\odot}}}M\frac{dn}{dM}(M)dM},
\end{equation}
where $\frac{dn}{dM}(M)$ is the mass function of dark matter haloes for late-type galaxies.  In this case only the slope of the halo mass function matters, and we assume this to be independent of the morphology of the central galaxy.

In Figure~\ref{barycen}, we use these formulae to plot the fractional abundances of the three baryon phases as a function of halo mass.  For this calculation, we have used the halo mass function for WMAP3 cosmology\footnote{This mass function was computed with HMFcalc \citep{Murray2013}.} \citep{Spergel2007}.  Over the entire mass range, $\sim85\%$ of the gas is expected to be in the hot phase, $\sim7-8\%$ of the total gas is in the cold phase, and $\sim6-7\%$ in stars. At least over this mass range, galaxy formation is therefore extremely inefficient: only $\sim13-15\%$ of the baryons exist as cold gas or stars.

As the mass of the halo increases, so does the fraction of the total baryons in stars.  For the DC14 models, the fraction of cold gas remains relatively constant as a function of halo mass and has a peak of $\sim8\%$ at ${\rm M_{halo}\sim6-8\times10^{10}\ M_{\odot}}$.  The NFW model exhibits a similar amplitude peak for the cold gas mass fraction, although this occurs at a marginally higher halo mass.  These curves may be used to constrain feedback models in simulations: not only should such models produce the right amount of stars, they should also give the correct proportion of cold gas. Although one would expect a significant amount of galaxy-to-galaxy scatter (visible in Figures~\ref{fd} and \ref{fs}), these baryon phase relations should be obtained on average over the galaxy population of a cosmological box.

\subsection{Caveats}
\label{caveats}

Star formation happens throughout the galaxy, and our assumption that all gas needs to be kicked from $r=0$ means that we overestimate to some extent the energy needed to eject the gas completely from the halo.  Furthermore, simply heating the gas to the virial temperature via SN is probably not enough to prevent all further star formation: the gas will buoyantly rise and then rain back down onto the disk as it cools. The specifics of this gas cycle can only really be teased out from numerical simulations that include all the relevant processes; however, results are still dependant on subgrid models of stellar feedback and resolution.  Therefore there are still large uncertainties in the amount of gas that actually re-accretes onto a galaxy as not all simulations can include all of the potentially important physics such as magnetic fields, thermal conduction, and cosmic rays \citep{Armillotta2016}.  

Our model assumes that if there were no stellar feedback, the cosmic fraction of baryons would settle into a cold disk of gas and stars. This is not necessarily the case as environment, mergers, and a variety of other cosmic mechanisms can disrupt this process. Indeed, both hydrodynamical simulation and semi-analytic models \citep{Nelson, Mitchell2018} predict that a significant fraction of the baryons do not accrete in the first place, and observations indicate that some gas is shocked and remains hot. Since galaxies form hierarchically in $\Lambda$CDM, the actual amount of baryons that are needed to removed from the cold disk of gas and stars is almost certainly less that what we have assumed in this paper as it is easier to eject baryons from one of the less massive progenitors. This makes our estimate of the required energy conservative. These processes may also introduce a mass dependence to the required feedback energy that our model does not account for.

``Missing baryons" may also be explained by heating the gas to a temperature that is not currently accessible by observations. This is not necessarily given by $T_{\rm vir}$, so one could replace $T_{\rm vir}$ in Equation~\ref{Eheat} with a fixed temperature, changing the mass dependence in the equation.

We have completely neglected other forms of feedback related to star formation such as photoionization, radiation pressure, cosmic rays, or stellar winds.  It is known that the energy input from stars in the form of radiation is roughly two orders of magnitude higher than the mechanical input from SN (e.g.~\citealt{Vecchia,Hopkins2014}). Simulations have shown that this radiation or even earlier stellar feedback in the form of winds can help regulate galaxy growth \citep{Hopkins2014}. The scalings presented here will continue to hold as long as the feedback energy scales linearly with ${\rm M_*}$. If, however, any of these processes are related to the mass of the galaxy, the predicted slope for the ${\rm M_*/M_{halo}}$ relation will cease to be $5/3$.  As noted above, it may be that ``preventative feedback" may be more effective at lower masses which could decrease the slope of the relation.

We have assumed that once an NFW cusp has been destroyed by feedback it will not reform. If it did, energy would have to be supplied continuously to maintain the core, which would significantly increase the required feedback efficiency for a given star formation rate. It is possible for mergers to recreate a cusp, but it has been shown that a core can be recreated even after the merger \citep[e.g.][]{Tollet2016}.  Galaxies likely evolve through a cyclic pattern of having cores and cusps depending on their ratio of ${\rm M_*/M_{halo}}$ \citep{Tollet2016}. Thus, while cusps tend to form again in dark matter-only simulations \citep{Laporte}, this may not necessarily be the case when full hydrodynamics is considered.  In particular, it is during mergers that cusps tend to form again in N-body only simulations, but mergers also trigger star formation which drives the outflows that may oppose this transformation.

Finally, all of our conclusions are subject to any selection effects that may be present in the SPARC sample.  This sample is representative of the the overall galaxy population in most key properties~\citep{Lelli2016}, but is neither complete nor volume limited.  We only consider late-type galaxies, which may in part cause our derived slope for the  ${\rm M_*/M_{halo}}$ relation to systematically disagree with literature relations \citep{Moster2013,Behroozi2013}.  Performing these same experiments with data sets which are either more or less representative or compiled in a more or less homogeneous fashion may lead to different results.

Given these uncertainties, we do not intend our analysis to prove that SN energy couples efficiently to the gas. We have however demonstrated that were this the case, this energy would be sufficient to create cores and produce rotation curves that match the SPARC sample.  This is consistent with other results from the literature that present similar findings \citep{Katz2016,Santos2018,Tollet2016}.

\section{Conclusion}
We have calculated the fraction of ``missing baryons" in 147 galaxies from the SPARC database, using empirically estimated halo masses from rotation curve fits for two different halo models: NFW and DC14 (see \citealt{Katz2016}). Our main results are as follows:

\begin{itemize}

\item{} We confirm that galaxy formation is a globally inefficient process.  The observed fraction of baryons present in the cold disk is in general much smaller than the cosmic baryon fraction. This fraction scales weakly with the virial mass of the galaxy such that $f_d\propto{\rm M_{vir}^{0.16}}$, although there is significant scatter about this relation.  Likewise, the fraction of baryonic mass comprised of stars scales strongly with mass; the most massive galaxies in our sample are completely dominated by stars.  At ${\rm M_{vir}\sim10^{11.1}}$, roughly 50\% of the observed baryonic content is in cold gas and the other 50\% is in stars.

\item{} When comparing the energy required to heat the ``missing baryons" to the virial temperature versus eject them from the halo, we find that far less energy is required for the former.  For SN feedback to regulate the heating, $\lesssim10\%$ of the total SN energy available needs to couple to the gas (dependent on the stellar IMF).  We find that a constant coupling efficiency is sufficient to explain the ``missing baryons" for late-type spiral galaxies in the mass range $10^9<{\rm M_{vir}/M_{\odot}}<10^{12}$ when haloes are modelled with a DC14 profile. This then sets the slope of the ${\rm M_*/M_{vir}}$ relation to be slightly less than $5/3$ due to the weak scaling of $f_d$ with ${\rm M_{vir}}$. However, there is typically not enough SN energy to eject all of the missing baryons from the halo.

\item{} When comparing the energy required to restructure the halo from a primordial NFW profile to the DC14 form, we find that there is always more energy available from SN feedback than required for the transformation.  We find that there is no mass dependence in the SN coupling efficiency required to create cores and that the magnitude of this efficiency is very similar to what is required to heat the ``missing baryons".  In contrast, there is a mass dependency in the efficiency required to erase the halo transformation for systems that exhibit contraction. This may indicate that halo restructuring is dependent on quantities other than stellar mass such as halo mass, formation redshift, or star formation history.

\item{} We take a cosmic census of the baryons that should be associated with late-type galaxies, finding that $\sim86\%$ of the total baryonic content is likely to be in a hot halo surrounding the galaxy, and the remaining $14\%$ split equally between stars and gas. In order for our predictions to be confirmed, future X-ray or kinetic Sunyaev-Zeldovich observations will have to accurately measure the masses in hot gas surrounding these local galaxies. 

\end{itemize}

\section*{Acknowledgements}
We thank the anonymous referee for their suggestions that improved the manuscript.  H.K. thanks the Beecroft fellowship, the Nicholas Kurti Junior Fellowship, and Brasenose College. H.D. is supported by St John's College, Oxford. ADC acknowledges financial support from a Marie Sklodowska-Curie Individual Fellowship grant, H2020-MSCA-IF-2016, Grant agreement 748213, DIGESTIVO.

\bibliographystyle{mnras.bst}
\bibliography{main.bib}

\begin{thebibliography}{}
\makeatletter
\relax
\def\mn@urlcharsother{\let\do\@makeother \do\$\do\&\do\#\do\^\do\_\do\%\do\~}
\def\mn@doi{\begingroup\mn@urlcharsother \@ifnextchar [ {\mn@doi@}
  {\mn@doi@[]}}
\def\mn@doi@[#1]#2{\def\@tempa{#1}\ifx\@tempa\@empty \href
  {http://dx.doi.org/#2} {doi:#2}\else \href {http://dx.doi.org/#2} {#1}\fi
  \endgroup}
\def\mn@eprint#1#2{\mn@eprint@#1:#2::\@nil}
\def\mn@eprint@arXiv#1{\href {http://arxiv.org/abs/#1} {{\tt arXiv:#1}}}
\def\mn@eprint@dblp#1{\href {http://dblp.uni-trier.de/rec/bibtex/#1.xml}
  {dblp:#1}}
\def\mn@eprint@#1:#2:#3:#4\@nil{\def\@tempa {#1}\def\@tempb {#2}\def\@tempc
  {#3}\ifx \@tempc \@empty \let \@tempc \@tempb \let \@tempb \@tempa \fi \ifx
  \@tempb \@empty \def\@tempb {arXiv}\fi \@ifundefined
  {mn@eprint@\@tempb}{\@tempb:\@tempc}{\expandafter \expandafter \csname
  mn@eprint@\@tempb\endcsname \expandafter{\@tempc}}}

\bibitem[\protect\citeauthoryear{{Anderson} \& {Bregman}}{{Anderson} \&
  {Bregman}}{2011}]{Anderson2011}
{Anderson} M.~E.,  {Bregman} J.~N.,  2011, \mn@doi [\apj]
  {10.1088/0004-637X/737/1/22}, \href
  {http://adsabs.harvard.edu/abs/2011ApJ...737...22A} {737, 22}

\bibitem[\protect\citeauthoryear{{Anderson}, {Bregman}  \& {Dai}}{{Anderson}
  et~al.}{2013}]{Anderson2013}
{Anderson} M.~E.,  {Bregman} J.~N.,   {Dai} X.,  2013, \mn@doi [\apj]
  {10.1088/0004-637X/762/2/106}, \href
  {http://adsabs.harvard.edu/abs/2013ApJ...762..106A} {762, 106}

\bibitem[\protect\citeauthoryear{{Armillotta}, {Fraternali}  \&
  {Marinacci}}{{Armillotta} et~al.}{2016}]{Armillotta2016}
{Armillotta} L.,  {Fraternali} F.,   {Marinacci} F.,  2016, \mn@doi [\mnras]
  {10.1093/mnras/stw1930}, \href
  {http://adsabs.harvard.edu/abs/2016MNRAS.462.4157A} {462, 4157}

\bibitem[\protect\citeauthoryear{{Babul} \& {Rees}}{{Babul} \&
  {Rees}}{1992}]{Babul1992}
{Babul} A.,  {Rees} M.~J.,  1992, \mn@doi [\mnras] {10.1093/mnras/255.2.346},
  \href {http://adsabs.harvard.edu/abs/1992MNRAS.255..346B} {255, 346}

\bibitem[\protect\citeauthoryear{{Behroozi}, {Wechsler}  \&
  {Conroy}}{{Behroozi} et~al.}{2013}]{Behroozi2013}
{Behroozi} P.~S.,  {Wechsler} R.~H.,   {Conroy} C.,  2013, \mn@doi [\apj]
  {10.1088/0004-637X/770/1/57}, \href
  {http://adsabs.harvard.edu/abs/2013ApJ...770...57B} {770, 57}

\bibitem[\protect\citeauthoryear{{Bernardi}, {Meert}, {Sheth}, {Vikram},
  {Huertas-Company}, {Mei}  \& {Shankar}}{{Bernardi}
  et~al.}{2013}]{Bernardi_SMF}
{Bernardi} M.,  {Meert} A.,  {Sheth} R.~K.,  {Vikram} V.,  {Huertas-Company}
  M.,  {Mei} S.,   {Shankar} F.,  2013, \mn@doi [\mnras]
  {10.1093/mnras/stt1607}, \href
  {http://adsabs.harvard.edu/abs/2013MNRAS.436..697B} {436, 697}

\bibitem[\protect\citeauthoryear{{Blumenthal}, {Faber}, {Flores}  \&
  {Primack}}{{Blumenthal} et~al.}{1986}]{Blumenthal1986}
{Blumenthal} G.~R.,  {Faber} S.~M.,  {Flores} R.,   {Primack} J.~R.,  1986,
  \mn@doi [\apj] {10.1086/163867}, \href
  {http://adsabs.harvard.edu/abs/1986ApJ...301...27B} {301, 27}

\bibitem[\protect\citeauthoryear{{Bower}, {Benson}, {Malbon}, {Helly}, {Frenk},
  {Baugh}, {Cole}  \& {Lacey}}{{Bower} et~al.}{2006}]{Bower2006}
{Bower} R.~G.,  {Benson} A.~J.,  {Malbon} R.,  {Helly} J.~C.,  {Frenk} C.~S.,
  {Baugh} C.~M.,  {Cole} S.,   {Lacey} C.~G.,  2006, \mn@doi [\mnras]
  {10.1111/j.1365-2966.2006.10519.x}, \href
  {http://adsabs.harvard.edu/abs/2006MNRAS.370..645B} {370, 645}

\bibitem[\protect\citeauthoryear{{Bradford}, {Geha}  \& {Blanton}}{{Bradford}
  et~al.}{2015}]{Bradford2015}
{Bradford} J.~D.,  {Geha} M.~C.,   {Blanton} M.~R.,  2015, \mn@doi [\apj]
  {10.1088/0004-637X/809/2/146}, \href
  {http://adsabs.harvard.edu/abs/2015ApJ...809..146B} {809, 146}

\bibitem[\protect\citeauthoryear{{Bregman}, {Alves}, {Miller}  \&
  {Hodges-Kluck}}{{Bregman} et~al.}{2015}]{Bregman2015}
{Bregman} J.~N.,  {Alves} G.~C.,  {Miller} M.~J.,   {Hodges-Kluck} E.,  2015,
  \mn@doi [Journal of Astronomical Telescopes, Instruments, and Systems]
  {10.1117/1.JATIS.1.4.045003}, \href
  {http://adsabs.harvard.edu/abs/2015JATIS...1d5003B} {1, 045003}

\bibitem[\protect\citeauthoryear{{Bregman}, {Anderson}, {Miller},
  {Hodges-Kluck}, {Dai}, {Li}, {Li}  \& {Qu}}{{Bregman}
  et~al.}{2018}]{Bregman2018}
{Bregman} J.~N.,  {Anderson} M.~E.,  {Miller} M.~J.,  {Hodges-Kluck} E.,  {Dai}
  X.,  {Li} J.-T.,  {Li} Y.,   {Qu} Z.,  2018, preprint, \href
  {http://adsabs.harvard.edu/abs/2018arXiv180308963B} {} (\mn@eprint {arXiv}
  {1803.08963})

\bibitem[\protect\citeauthoryear{{Breitschwerdt}, {McKenzie}  \&
  {Voelk}}{{Breitschwerdt} et~al.}{1991}]{Breit1991}
{Breitschwerdt} D.,  {McKenzie} J.~F.,   {Voelk} H.~J.,  1991, \aap, \href
  {http://adsabs.harvard.edu/abs/1991A%26A...245...79B} {245, 79}

\bibitem[\protect\citeauthoryear{{Brook} \& {Di Cintio}}{{Brook} \& {Di
  Cintio}}{2015}]{Brook2015}
{Brook} C.~B.,  {Di Cintio} A.,  2015, \mn@doi [\mnras]
  {10.1093/mnras/stv1699}, \href
  {http://adsabs.harvard.edu/abs/2015MNRAS.453.2133B} {453, 2133}

\bibitem[\protect\citeauthoryear{{Brook}, {Stinson}, {Gibson}, {Wadsley}  \&
  {Quinn}}{{Brook} et~al.}{2012}]{Brook2012}
{Brook} C.~B.,  {Stinson} G.,  {Gibson} B.~K.,  {Wadsley} J.,   {Quinn} T.,
  2012, \mn@doi [\mnras] {10.1111/j.1365-2966.2012.21306.x}, \href
  {http://adsabs.harvard.edu/abs/2012MNRAS.424.1275B} {424, 1275}

\bibitem[\protect\citeauthoryear{{Brook}, {Stinson}, {Gibson}, {Shen},
  {Macci{\`o}}, {Obreja}, {Wadsley}  \& {Quinn}}{{Brook}
  et~al.}{2014}]{Brook2014}
{Brook} C.~B.,  {Stinson} G.,  {Gibson} B.~K.,  {Shen} S.,  {Macci{\`o}} A.~V.,
   {Obreja} A.,  {Wadsley} J.,   {Quinn} T.,  2014, \mn@doi [\mnras]
  {10.1093/mnras/stu1406}, \href
  {http://adsabs.harvard.edu/abs/2014MNRAS.443.3809B} {443, 3809}

\bibitem[\protect\citeauthoryear{{Cen} \& {Ostriker}}{{Cen} \&
  {Ostriker}}{1999}]{Cen1999}
{Cen} R.,  {Ostriker} J.~P.,  1999, \mn@doi [\apj] {10.1086/306949}, \href
  {http://adsabs.harvard.edu/abs/1999ApJ...514....1C} {514, 1}

\bibitem[\protect\citeauthoryear{{Chabrier}}{{Chabrier}}{2003}]{Chabrier2003}
{Chabrier} G.,  2003, \mn@doi [\pasp] {10.1086/376392}, \href
  {http://adsabs.harvard.edu/abs/2003PASP..115..763C} {115, 763}

\bibitem[\protect\citeauthoryear{{Chabrier}, {Baraffe}, {Allard}  \&
  {Hauschildt}}{{Chabrier} et~al.}{2005}]{Chabrier2005}
{Chabrier} G.,  {Baraffe} I.,  {Allard} F.,   {Hauschildt} P.~H.,  2005, ArXiv
  Astrophysics e-prints, \href
  {http://adsabs.harvard.edu/abs/2005astro.ph..9798C} {}

\bibitem[\protect\citeauthoryear{Chan, Kereš, Oñorbe, Hopkins, Muratov,
  Faucher-Giguère  \& Quataert}{Chan et~al.}{2015}]{Chan}
Chan T.~K.,  Kereš D.,  Oñorbe J.,  Hopkins P.~F.,  Muratov A.~L.,
  Faucher-Giguère C.~A.,   Quataert E.,  2015, \mn@doi [Mon. Not. Roy. Astron.
  Soc.] {10.1093/mnras/stv2165}, 454, 2981

\bibitem[\protect\citeauthoryear{{Christensen}, {Dav{\'e}}, {Governato},
  {Pontzen}, {Brooks}, {Munshi}, {Quinn}  \& {Wadsley}}{{Christensen}
  et~al.}{2016}]{Christensen2016}
{Christensen} C.~R.,  {Dav{\'e}} R.,  {Governato} F.,  {Pontzen} A.,  {Brooks}
  A.,  {Munshi} F.,  {Quinn} T.,   {Wadsley} J.,  2016, \mn@doi [\apj]
  {10.3847/0004-637X/824/1/57}, \href
  {http://adsabs.harvard.edu/abs/2016ApJ...824...57C} {824, 57}

\bibitem[\protect\citeauthoryear{{Concas}, {Popesso}, {Brusa}, {Mainieri},
  {Erfanianfar}  \& {Morselli}}{{Concas} et~al.}{2017}]{Concas}
{Concas} A.,  {Popesso} P.,  {Brusa} M.,  {Mainieri} V.,  {Erfanianfar} G.,
  {Morselli} L.,  2017, \mn@doi [\aap] {10.1051/0004-6361/201629519}, \href
  {http://adsabs.harvard.edu/abs/2017A%26A...606A..36C} {606, A36}

\bibitem[\protect\citeauthoryear{{Costa}, {Sijacki}  \& {Haehnelt}}{{Costa}
  et~al.}{2014}]{Costa2014}
{Costa} T.,  {Sijacki} D.,   {Haehnelt} M.~G.,  2014, \mn@doi [\mnras]
  {10.1093/mnras/stu1632}, \href
  {http://adsabs.harvard.edu/abs/2014MNRAS.444.2355C} {444, 2355}

\bibitem[\protect\citeauthoryear{{Cowie}, {McKee}  \& {Ostriker}}{{Cowie}
  et~al.}{1981}]{Cowie1981}
{Cowie} L.~L.,  {McKee} C.~F.,   {Ostriker} J.~P.,  1981, \mn@doi [\apj]
  {10.1086/159100}, \href {http://adsabs.harvard.edu/abs/1981ApJ...247..908C}
  {247, 908}

\bibitem[\protect\citeauthoryear{{Curtis} \& {Sijacki}}{{Curtis} \&
  {Sijacki}}{2016}]{Curtis2016}
{Curtis} M.,  {Sijacki} D.,  2016, \mn@doi [\mnras] {10.1093/mnras/stw1944},
  \href {http://adsabs.harvard.edu/abs/2016MNRAS.tmp.1066C} {}

\bibitem[\protect\citeauthoryear{Dalla~Vecchia \& Schaye}{Dalla~Vecchia \&
  Schaye}{2008}]{Vecchia}
Dalla~Vecchia C.,  Schaye J.,  2008, \mn@doi [Monthly Notices of the Royal
  Astronomical Society] {10.1111/j.1365-2966.2008.13322.x}, 387, 1431

\bibitem[\protect\citeauthoryear{{Danforth} \& {Shull}}{{Danforth} \&
  {Shull}}{2005}]{Danforth2005}
{Danforth} C.~W.,  {Shull} J.~M.,  2005, \mn@doi [\apj] {10.1086/429285}, \href
  {http://adsabs.harvard.edu/abs/2005ApJ...624..555D} {624, 555}

\bibitem[\protect\citeauthoryear{{Dekel} \& {Silk}}{{Dekel} \&
  {Silk}}{1986}]{Dekel1986}
{Dekel} A.,  {Silk} J.,  1986, \mn@doi [\apj] {10.1086/164050}, \href
  {http://adsabs.harvard.edu/abs/1986ApJ...303...39D} {303, 39}

\bibitem[\protect\citeauthoryear{{Dekel}, {Ishai}, {Dutton}  \&
  {Maccio}}{{Dekel} et~al.}{2017}]{Dekel2017}
{Dekel} A.,  {Ishai} G.,  {Dutton} A.~A.,   {Maccio} A.~V.,  2017, \mn@doi
  [\mnras] {10.1093/mnras/stx486}, \href
  {http://adsabs.harvard.edu/abs/2017MNRAS.468.1005D} {468, 1005}

\bibitem[\protect\citeauthoryear{{Di Cintio}, {Brook}, {Macci{\`o}}, {Stinson},
  {Knebe}, {Dutton}  \& {Wadsley}}{{Di Cintio} et~al.}{2014a}]{DiCintio2014b}
{Di Cintio} A.,  {Brook} C.~B.,  {Macci{\`o}} A.~V.,  {Stinson} G.~S.,  {Knebe}
  A.,  {Dutton} A.~A.,   {Wadsley} J.,  2014a, \mn@doi [\mnras]
  {10.1093/mnras/stt1891}, \href
  {http://adsabs.harvard.edu/abs/2014MNRAS.437..415D} {437, 415}

\bibitem[\protect\citeauthoryear{{Di Cintio}, {Brook}, {Dutton}, {Macci{\`o}},
  {Stinson}  \& {Knebe}}{{Di Cintio} et~al.}{2014b}]{DC2014}
{Di Cintio} A.,  {Brook} C.~B.,  {Dutton} A.~A.,  {Macci{\`o}} A.~V.,
  {Stinson} G.~S.,   {Knebe} A.,  2014b, \mn@doi [\mnras]
  {10.1093/mnras/stu729}, \href
  {http://adsabs.harvard.edu/abs/2014MNRAS.441.2986D} {441, 2986}

\bibitem[\protect\citeauthoryear{{Di Cintio}, {Tremmel}, {Governato},
  {Pontzen}, {Zavala}, {Bastidas Fry}, {Brooks}  \& {Vogelsberger}}{{Di Cintio}
  et~al.}{2017}]{DiCintio2017}
{Di Cintio} A.,  {Tremmel} M.,  {Governato} F.,  {Pontzen} A.,  {Zavala} J.,
  {Bastidas Fry} A.,  {Brooks} A.,   {Vogelsberger} M.,  2017, \mn@doi [\mnras]
  {10.1093/mnras/stx1043}, \href
  {http://adsabs.harvard.edu/abs/2017MNRAS.469.2845D} {469, 2845}

\bibitem[\protect\citeauthoryear{{Dubois} et~al.,}{{Dubois}
  et~al.}{2014}]{Dubois2014}
{Dubois} Y.,  et~al., 2014, \mn@doi [\mnras] {10.1093/mnras/stu1227}, \href
  {http://adsabs.harvard.edu/abs/2014MNRAS.444.1453D} {444, 1453}

\bibitem[\protect\citeauthoryear{{Dutton} \& {Macci{\`o}}}{{Dutton} \&
  {Macci{\`o}}}{2014}]{Dutton2014}
{Dutton} A.~A.,  {Macci{\`o}} A.~V.,  2014, \mn@doi [\mnras]
  {10.1093/mnras/stu742}, \href
  {http://adsabs.harvard.edu/abs/2014MNRAS.441.3359D} {441, 3359}

\bibitem[\protect\citeauthoryear{{Dutton} et~al.,}{{Dutton}
  et~al.}{2011}]{Dutton_2011}
{Dutton} A.~A.,  et~al., 2011, \mn@doi [\mnras]
  {10.1111/j.1365-2966.2011.19038.x}, \href
  {http://adsabs.harvard.edu/abs/2011MNRAS.416..322D} {416, 322}

\bibitem[\protect\citeauthoryear{{Efstathiou}}{{Efstathiou}}{1992}]{Efstathiou1992}
{Efstathiou} G.,  1992, \mn@doi [\mnras] {10.1093/mnras/256.1.43P}, \href
  {http://adsabs.harvard.edu/abs/1992MNRAS.256P..43E} {256, 43P}

\bibitem[\protect\citeauthoryear{{Efstathiou}}{{Efstathiou}}{2000}]{Efstathiou}
{Efstathiou} G.,  2000, \mn@doi [\mnras] {10.1046/j.1365-8711.2000.03665.x},
  \href {http://adsabs.harvard.edu/abs/2000MNRAS.317..697E} {317, 697}

\bibitem[\protect\citeauthoryear{{Elbaz} et~al.,}{{Elbaz}
  et~al.}{2007}]{Elbaz2007}
{Elbaz} D.,  et~al., 2007, \mn@doi [\aap] {10.1051/0004-6361:20077525}, \href
  {http://adsabs.harvard.edu/abs/2007A%26A...468...33E} {468, 33}

\bibitem[\protect\citeauthoryear{{Fall} \& {Efstathiou}}{{Fall} \&
  {Efstathiou}}{1980}]{Fall1980}
{Fall} S.~M.,  {Efstathiou} G.,  1980, \mn@doi [\mnras]
  {10.1093/mnras/193.2.189}, \href
  {http://adsabs.harvard.edu/abs/1980MNRAS.193..189F} {193, 189}

\bibitem[\protect\citeauthoryear{Fischler \& Bolles}{Fischler \&
  Bolles}{1981}]{Fischler1981}
Fischler M.~A.,  Bolles R.~C.,  1981, \mn@doi [Commun. ACM]
  {10.1145/358669.358692}, 24, 381

\bibitem[\protect\citeauthoryear{{Fukugita}, {Hogan}  \& {Peebles}}{{Fukugita}
  et~al.}{1998}]{Fukugita1998}
{Fukugita} M.,  {Hogan} C.~J.,   {Peebles} P.~J.~E.,  1998, \mn@doi [\apj]
  {10.1086/306025}, \href {http://adsabs.harvard.edu/abs/1998ApJ...503..518F}
  {503, 518}

\bibitem[\protect\citeauthoryear{{Gentry}, {Krumholz}, {Dekel}  \&
  {Madau}}{{Gentry} et~al.}{2017}]{Gentry2017}
{Gentry} E.~S.,  {Krumholz} M.~R.,  {Dekel} A.,   {Madau} P.,  2017, \mn@doi
  [\mnras] {10.1093/mnras/stw2746}, \href
  {http://adsabs.harvard.edu/abs/2017MNRAS.465.2471G} {465, 2471}

\bibitem[\protect\citeauthoryear{{Giodini} et~al.,}{{Giodini}
  et~al.}{2009}]{Giodini2009}
{Giodini} S.,  et~al., 2009, \mn@doi [\apj] {10.1088/0004-637X/703/1/982},
  \href {http://adsabs.harvard.edu/abs/2009ApJ...703..982G} {703, 982}

\bibitem[\protect\citeauthoryear{{Gnedin}}{{Gnedin}}{2000}]{Gnedin2000}
{Gnedin} N.~Y.,  2000, \mn@doi [\apj] {10.1086/317042}, \href
  {http://adsabs.harvard.edu/abs/2000ApJ...542..535G} {542, 535}

\bibitem[\protect\citeauthoryear{{Gnedin}, {Kravtsov}, {Klypin}  \&
  {Nagai}}{{Gnedin} et~al.}{2004}]{Gnedin2004}
{Gnedin} O.~Y.,  {Kravtsov} A.~V.,  {Klypin} A.~A.,   {Nagai} D.,  2004,
  \mn@doi [\apj] {10.1086/424914}, \href
  {http://adsabs.harvard.edu/abs/2004ApJ...616...16G} {616, 16}

\bibitem[\protect\citeauthoryear{{Gonzalez}, {Sivanandam}, {Zabludoff}  \&
  {Zaritsky}}{{Gonzalez} et~al.}{2013}]{Gonzalez2013}
{Gonzalez} A.~H.,  {Sivanandam} S.,  {Zabludoff} A.~I.,   {Zaritsky} D.,  2013,
  \mn@doi [\apj] {10.1088/0004-637X/778/1/14}, \href
  {http://adsabs.harvard.edu/abs/2013ApJ...778...14G} {778, 14}

\bibitem[\protect\citeauthoryear{{Hernquist}}{{Hernquist}}{1990}]{Hernquist1990}
{Hernquist} L.,  1990, \mn@doi [\apj] {10.1086/168845}, \href
  {http://adsabs.harvard.edu/abs/1990ApJ...356..359H} {356, 359}

\bibitem[\protect\citeauthoryear{{Hopkins}, {Kere{\v s}}, {Murray}, {Quataert}
  \& {Hernquist}}{{Hopkins} et~al.}{2012}]{Hopkins2012}
{Hopkins} P.~F.,  {Kere{\v s}} D.,  {Murray} N.,  {Quataert} E.,   {Hernquist}
  L.,  2012, \mn@doi [\mnras] {10.1111/j.1365-2966.2012.21981.x}, \href
  {http://adsabs.harvard.edu/abs/2012MNRAS.427..968H} {427, 968}

\bibitem[\protect\citeauthoryear{{Hopkins}, {Kere{\v s}}, {O{\~n}orbe},
  {Faucher-Gigu{\`e}re}, {Quataert}, {Murray}  \& {Bullock}}{{Hopkins}
  et~al.}{2014}]{Hopkins2014}
{Hopkins} P.~F.,  {Kere{\v s}} D.,  {O{\~n}orbe} J.,  {Faucher-Gigu{\`e}re}
  C.-A.,  {Quataert} E.,  {Murray} N.,   {Bullock} J.~S.,  2014, \mn@doi
  [\mnras] {10.1093/mnras/stu1738}, \href
  {http://adsabs.harvard.edu/abs/2014MNRAS.445..581H} {445, 581}

\bibitem[\protect\citeauthoryear{{Hopkins} et~al.,}{{Hopkins}
  et~al.}{2017}]{Hopkins2017}
{Hopkins} P.~F.,  et~al., 2017, preprint, \href
  {http://adsabs.harvard.edu/abs/2017arXiv170206148H} {} (\mn@eprint {arXiv}
  {1702.06148})

\bibitem[\protect\citeauthoryear{{Karim} et~al.,}{{Karim}
  et~al.}{2011}]{Karim2011}
{Karim} A.,  et~al., 2011, \mn@doi [\apj] {10.1088/0004-637X/730/2/61}, \href
  {http://adsabs.harvard.edu/abs/2011ApJ...730...61K} {730, 61}

\bibitem[\protect\citeauthoryear{{Katz}, {McGaugh}, {Sellwood}  \& {de
  Blok}}{{Katz} et~al.}{2014}]{Katz2014}
{Katz} H.,  {McGaugh} S.~S.,  {Sellwood} J.~A.,   {de Blok} W.~J.~G.,  2014,
  \mn@doi [\mnras] {10.1093/mnras/stu070}, \href
  {http://adsabs.harvard.edu/abs/2014MNRAS.439.1897K} {439, 1897}

\bibitem[\protect\citeauthoryear{{Katz}, {Lelli}, {McGaugh}, {Di Cintio},
  {Brook}  \& {Schombert}}{{Katz} et~al.}{2017}]{Katz2016}
{Katz} H.,  {Lelli} F.,  {McGaugh} S.~S.,  {Di Cintio} A.,  {Brook} C.~B.,
  {Schombert} J.~M.,  2017, \mn@doi [\mnras] {10.1093/mnras/stw3101}, \href
  {http://adsabs.harvard.edu/abs/2017MNRAS.466.1648K} {466, 1648}

\bibitem[\protect\citeauthoryear{{Kere{\v s}}, {Katz}, {Weinberg}  \&
  {Dav{\'e}}}{{Kere{\v s}} et~al.}{2005}]{Keres2005}
{Kere{\v s}} D.,  {Katz} N.,  {Weinberg} D.~H.,   {Dav{\'e}} R.,  2005, \mn@doi
  [\mnras] {10.1111/j.1365-2966.2005.09451.x}, \href
  {http://adsabs.harvard.edu/abs/2005MNRAS.363....2K} {363, 2}

\bibitem[\protect\citeauthoryear{{King}}{{King}}{2003}]{King2003}
{King} A.,  2003, \mn@doi [\apjl] {10.1086/379143}, \href
  {http://adsabs.harvard.edu/abs/2003ApJ...596L..27K} {596, L27}

\bibitem[\protect\citeauthoryear{{Laporte} \& {Pe{\~n}arrubia}}{{Laporte} \&
  {Pe{\~n}arrubia}}{2015}]{Laporte}
{Laporte} C.~F.~P.,  {Pe{\~n}arrubia} J.,  2015, \mn@doi [\mnras]
  {10.1093/mnrasl/slv008}, \href
  {http://adsabs.harvard.edu/abs/2015MNRAS.449L..90L} {449, L90}

\bibitem[\protect\citeauthoryear{{Lelli}, {Verheijen}  \& {Fraternali}}{{Lelli}
  et~al.}{2014}]{Lelli_starburst}
{Lelli} F.,  {Verheijen} M.,   {Fraternali} F.,  2014, \mn@doi [\aap]
  {10.1051/0004-6361/201322657}, \href
  {http://adsabs.harvard.edu/abs/2014A%26A...566A..71L} {566, A71}

\bibitem[\protect\citeauthoryear{{Lelli}, {McGaugh}  \& {Schombert}}{{Lelli}
  et~al.}{2016}]{Lelli2016}
{Lelli} F.,  {McGaugh} S.~S.,   {Schombert} J.~M.,  2016, preprint, \href
  {http://adsabs.harvard.edu/abs/2016arXiv160609251L} {} (\mn@eprint {arXiv}
  {1606.09251})

\bibitem[\protect\citeauthoryear{{Li} \& {White}}{{Li} \&
  {White}}{2009}]{Li2009}
{Li} C.,  {White} S.~D.~M.,  2009, \mn@doi [\mnras]
  {10.1111/j.1365-2966.2009.15268.x}, \href
  {http://adsabs.harvard.edu/abs/2009MNRAS.398.2177L} {398, 2177}

\bibitem[\protect\citeauthoryear{{Li}, {Bregman}, {Wang}, {Crain}  \&
  {Anderson}}{{Li} et~al.}{2018}]{Li2018}
{Li} J.-T.,  {Bregman} J.~N.,  {Wang} Q.~D.,  {Crain} R.~A.,   {Anderson}
  M.~E.,  2018, preprint, \href
  {http://adsabs.harvard.edu/abs/2018arXiv180209453L} {} (\mn@eprint {arXiv}
  {1802.09453})

\bibitem[\protect\citeauthoryear{{Macci{\`o}}, {Dutton}  \& {van den
  Bosch}}{{Macci{\`o}} et~al.}{2008}]{Maccio2008}
{Macci{\`o}} A.~V.,  {Dutton} A.~A.,   {van den Bosch} F.~C.,  2008, \mn@doi
  [\mnras] {10.1111/j.1365-2966.2008.14029.x}, \href
  {http://adsabs.harvard.edu/abs/2008MNRAS.391.1940M} {391, 1940}

\bibitem[\protect\citeauthoryear{{Marinacci}, {Binney}, {Fraternali}, {Nipoti},
  {Ciotti}  \& {Londrillo}}{{Marinacci} et~al.}{2010}]{Marinacci}
{Marinacci} F.,  {Binney} J.,  {Fraternali} F.,  {Nipoti} C.,  {Ciotti} L.,
  {Londrillo} P.,  2010, \mn@doi [\mnras] {10.1111/j.1365-2966.2010.16352.x},
  \href {http://adsabs.harvard.edu/abs/2010MNRAS.404.1464M} {404, 1464}

\bibitem[\protect\citeauthoryear{{Martizzi}, {Teyssier}  \& {Moore}}{{Martizzi}
  et~al.}{2013}]{Martizzi2013}
{Martizzi} D.,  {Teyssier} R.,   {Moore} B.,  2013, \mn@doi [\mnras]
  {10.1093/mnras/stt297}, \href
  {http://adsabs.harvard.edu/abs/2013MNRAS.432.1947M} {432, 1947}

\bibitem[\protect\citeauthoryear{{Mathews} \& {Baker}}{{Mathews} \&
  {Baker}}{1971}]{Mathews1971}
{Mathews} W.~G.,  {Baker} J.~C.,  1971, \mn@doi [\apj] {10.1086/151208}, \href
  {http://adsabs.harvard.edu/abs/1971ApJ...170..241M} {170, 241}

\bibitem[\protect\citeauthoryear{{Maxwell}, {Wadsley}  \& {Couchman}}{{Maxwell}
  et~al.}{2015}]{Maxwell2015}
{Maxwell} A.~J.,  {Wadsley} J.,   {Couchman} H.~M.~P.,  2015, \mn@doi [\apj]
  {10.1088/0004-637X/806/2/229}, \href
  {http://adsabs.harvard.edu/abs/2015ApJ...806..229M} {806, 229}

\bibitem[\protect\citeauthoryear{{McGaugh} \& {Schombert}}{{McGaugh} \&
  {Schombert}}{2014}]{Mcgaugh2014}
{McGaugh} S.~S.,  {Schombert} J.~M.,  2014, \mn@doi [\aj]
  {10.1088/0004-6256/148/5/77}, \href
  {http://adsabs.harvard.edu/abs/2014AJ....148...77M} {148, 77}

\bibitem[\protect\citeauthoryear{{McGaugh}, {Schombert}, {de Blok}  \&
  {Zagursky}}{{McGaugh} et~al.}{2010}]{McGaugh2010}
{McGaugh} S.~S.,  {Schombert} J.~M.,  {de Blok} W.~J.~G.,   {Zagursky} M.~J.,
  2010, \mn@doi [\apjl] {10.1088/2041-8205/708/1/L14}, \href
  {http://adsabs.harvard.edu/abs/2010ApJ...708L..14M} {708, L14}

\bibitem[\protect\citeauthoryear{{McKee} \& {Ostriker}}{{McKee} \&
  {Ostriker}}{1977}]{McKee1977}
{McKee} C.~F.,  {Ostriker} J.~P.,  1977, \mn@doi [\apj] {10.1086/155667}, \href
  {http://adsabs.harvard.edu/abs/1977ApJ...218..148M} {218, 148}

\bibitem[\protect\citeauthoryear{{Miller} \& {Bregman}}{{Miller} \&
  {Bregman}}{2015}]{Miller2015}
{Miller} M.~J.,  {Bregman} J.~N.,  2015, \mn@doi [\apj]
  {10.1088/0004-637X/800/1/14}, \href
  {http://adsabs.harvard.edu/abs/2015ApJ...800...14M} {800, 14}

\bibitem[\protect\citeauthoryear{{Mitchell}, {Blaizot}, {Devriendt}, {Kimm},
  {Michel-Dansac}, {Rosdahl}  \& {Slyz}}{{Mitchell}
  et~al.}{2017}]{Mitchell2017}
{Mitchell} P.,  {Blaizot} J.,  {Devriendt} J.,  {Kimm} T.,  {Michel-Dansac} L.,
   {Rosdahl} J.,   {Slyz} A.,  2017, preprint, \href
  {http://adsabs.harvard.edu/abs/2017arXiv171003765M} {} (\mn@eprint {arXiv}
  {1710.03765})

\bibitem[\protect\citeauthoryear{{Mitchell} et~al.,}{{Mitchell}
  et~al.}{2018}]{Mitchell2018}
{Mitchell} P.~D.,  et~al., 2018, \mn@doi [\mnras] {10.1093/mnras/stx2770},
  \href {http://adsabs.harvard.edu/abs/2018MNRAS.474..492M} {474, 492}

\bibitem[\protect\citeauthoryear{{Moster}, {Naab}  \& {White}}{{Moster}
  et~al.}{2013}]{Moster2013}
{Moster} B.~P.,  {Naab} T.,   {White} S.~D.~M.,  2013, \mn@doi [\mnras]
  {10.1093/mnras/sts261}, \href
  {http://adsabs.harvard.edu/abs/2013MNRAS.428.3121M} {428, 3121}

\bibitem[\protect\citeauthoryear{{Moster}, {Naab}  \& {White}}{{Moster}
  et~al.}{2017}]{Moster2017}
{Moster} B.~P.,  {Naab} T.,   {White} S.~D.~M.,  2017, preprint, \href
  {http://adsabs.harvard.edu/abs/2017arXiv170505373M} {} (\mn@eprint {arXiv}
  {1705.05373})

\bibitem[\protect\citeauthoryear{{Murray}, {Power}  \& {Robotham}}{{Murray}
  et~al.}{2013}]{Murray2013}
{Murray} S.~G.,  {Power} C.,   {Robotham} A.~S.~G.,  2013, \mn@doi [Astronomy
  and Computing] {10.1016/j.ascom.2013.11.001}, \href
  {http://adsabs.harvard.edu/abs/2013A%26C.....3...23M} {3, 23}

\bibitem[\protect\citeauthoryear{{Navarro}, {Eke}  \& {Frenk}}{{Navarro}
  et~al.}{1996a}]{Navarro1996b}
{Navarro} J.~F.,  {Eke} V.~R.,   {Frenk} C.~S.,  1996a, \mn@doi [\mnras]
  {10.1093/mnras/283.3.L72}, \href
  {http://adsabs.harvard.edu/abs/1996MNRAS.283L..72N} {283, L72}

\bibitem[\protect\citeauthoryear{{Navarro}, {Frenk}  \& {White}}{{Navarro}
  et~al.}{1996b}]{Navarro1996}
{Navarro} J.~F.,  {Frenk} C.~S.,   {White} S.~D.~M.,  1996b, \mn@doi [\apj]
  {10.1086/177173}, \href {http://adsabs.harvard.edu/abs/1996ApJ...462..563N}
  {462, 563}

\bibitem[\protect\citeauthoryear{{Nelson}, {Genel}, {Pillepich},
  {Vogelsberger}, {Springel}  \& {Hernquist}}{{Nelson} et~al.}{2016}]{Nelson}
{Nelson} D.,  {Genel} S.,  {Pillepich} A.,  {Vogelsberger} M.,  {Springel} V.,
   {Hernquist} L.,  2016, \mn@doi [\mnras] {10.1093/mnras/stw1191}, \href
  {http://adsabs.harvard.edu/abs/2016MNRAS.460.2881N} {460, 2881}

\bibitem[\protect\citeauthoryear{{Okamoto}, {Gao}  \& {Theuns}}{{Okamoto}
  et~al.}{2008}]{Okamoto2008}
{Okamoto} T.,  {Gao} L.,   {Theuns} T.,  2008, \mn@doi [\mnras]
  {10.1111/j.1365-2966.2008.13830.x}, \href
  {http://adsabs.harvard.edu/abs/2008MNRAS.390..920O} {390, 920}

\bibitem[\protect\citeauthoryear{{Papastergis}, {Cattaneo}, {Huang},
  {Giovanelli}  \& {Haynes}}{{Papastergis} et~al.}{2012}]{Papastergis2013}
{Papastergis} E.,  {Cattaneo} A.,  {Huang} S.,  {Giovanelli} R.,   {Haynes}
  M.~P.,  2012, \mn@doi [\apj] {10.1088/0004-637X/759/2/138}, \href
  {http://adsabs.harvard.edu/abs/2012ApJ...759..138P} {759, 138}

\bibitem[\protect\citeauthoryear{{Pe{\~n}arrubia}, {Pontzen}, {Walker}  \&
  {Koposov}}{{Pe{\~n}arrubia} et~al.}{2012}]{Penarrubia2012}
{Pe{\~n}arrubia} J.,  {Pontzen} A.,  {Walker} M.~G.,   {Koposov} S.~E.,  2012,
  \mn@doi [\apjl] {10.1088/2041-8205/759/2/L42}, \href
  {http://adsabs.harvard.edu/abs/2012ApJ...759L..42P} {759, L42}

\bibitem[\protect\citeauthoryear{Pedregosa et~al.,}{Pedregosa
  et~al.}{2011}]{sklearn}
Pedregosa F.,  et~al., 2011, Journal of Machine Learning Research, 12, 2825

\bibitem[\protect\citeauthoryear{{Planck Collaboration} et~al.,}{{Planck
  Collaboration} et~al.}{2015}]{Planck2015}
{Planck Collaboration} et~al., 2015, preprint, \href
  {http://adsabs.harvard.edu/abs/2015arXiv150201589P} {} (\mn@eprint {arXiv}
  {1502.01589})

\bibitem[\protect\citeauthoryear{{Pontzen} \& {Governato}}{{Pontzen} \&
  {Governato}}{2012}]{Pontzen2012}
{Pontzen} A.,  {Governato} F.,  2012, \mn@doi [\mnras]
  {10.1111/j.1365-2966.2012.20571.x}, \href
  {http://adsabs.harvard.edu/abs/2012MNRAS.421.3464P} {421, 3464}

\bibitem[\protect\citeauthoryear{{Puchwein}, {Sijacki}  \&
  {Springel}}{{Puchwein} et~al.}{2008}]{Puchwein2008}
{Puchwein} E.,  {Sijacki} D.,   {Springel} V.,  2008, \mn@doi [\apjl]
  {10.1086/593352}, \href {http://adsabs.harvard.edu/abs/2008ApJ...687L..53P}
  {687, L53}

\bibitem[\protect\citeauthoryear{{Read} \& {Gilmore}}{{Read} \&
  {Gilmore}}{2005}]{Read2005}
{Read} J.~I.,  {Gilmore} G.,  2005, \mn@doi [\mnras]
  {10.1111/j.1365-2966.2004.08424.x}, \href
  {http://adsabs.harvard.edu/abs/2005MNRAS.356..107R} {356, 107}

\bibitem[\protect\citeauthoryear{{Read}, {Agertz}  \& {Collins}}{{Read}
  et~al.}{2016a}]{Read2016}
{Read} J.~I.,  {Agertz} O.,   {Collins} M.~L.~M.,  2016a, \mn@doi [\mnras]
  {10.1093/mnras/stw713}, \href
  {http://adsabs.harvard.edu/abs/2016MNRAS.459.2573R} {459, 2573}

\bibitem[\protect\citeauthoryear{Read, Agertz  \& Collins}{Read
  et~al.}{2016b}]{Read}
Read J.~I.,  Agertz O.,   Collins M. L.~M.,  2016b, \mn@doi [Mon. Not. Roy.
  Astron. Soc.] {10.1093/mnras/stw713}, 459, 2573

\bibitem[\protect\citeauthoryear{{Rogers} \& {Pittard}}{{Rogers} \&
  {Pittard}}{2013}]{Rogers2013}
{Rogers} H.,  {Pittard} J.~M.,  2013, \mn@doi [\mnras] {10.1093/mnras/stt255},
  \href {http://adsabs.harvard.edu/abs/2013MNRAS.431.1337R} {431, 1337}

\bibitem[\protect\citeauthoryear{{Salpeter}}{{Salpeter}}{1955}]{Salpeter1955}
{Salpeter} E.~E.,  1955, \mn@doi [\apj] {10.1086/145971}, \href
  {http://adsabs.harvard.edu/abs/1955ApJ...121..161S} {121, 161}

\bibitem[\protect\citeauthoryear{{Santos-Santos}, {Di Cintio}, {Brook},
  {Macci{\`o}}, {Dutton}  \& {Dom{\'{\i}}nguez-Tenreiro}}{{Santos-Santos}
  et~al.}{2018}]{Santos2018}
{Santos-Santos} I.~M.,  {Di Cintio} A.,  {Brook} C.~B.,  {Macci{\`o}} A.,
  {Dutton} A.,   {Dom{\'{\i}}nguez-Tenreiro} R.,  2018, \mn@doi [\mnras]
  {10.1093/mnras/stx2660}, \href
  {http://adsabs.harvard.edu/abs/2018MNRAS.473.4392S} {473, 4392}

\bibitem[\protect\citeauthoryear{Sawala et~al.}{Sawala et~al.}{2016}]{Apostle}
Sawala T.,  et~al., 2016, \mn@doi [Mon. Not. Roy. Astron. Soc.]
  {10.1093/mnras/stw145}, 457, 1931

\bibitem[\protect\citeauthoryear{{Schaye} et~al.,}{{Schaye}
  et~al.}{2015}]{Schaye2015}
{Schaye} J.,  et~al., 2015, \mn@doi [\mnras] {10.1093/mnras/stu2058}, \href
  {http://adsabs.harvard.edu/abs/2015MNRAS.446..521S} {446, 521}

\bibitem[\protect\citeauthoryear{{Sedov}}{{Sedov}}{1959}]{Sedov1959}
{Sedov} L.~I.,  1959, {Similarity and Dimensional Methods in Mechanics}

\bibitem[\protect\citeauthoryear{{Shull}, {Smith}  \& {Danforth}}{{Shull}
  et~al.}{2012}]{Shull2012}
{Shull} J.~M.,  {Smith} B.~D.,   {Danforth} C.~W.,  2012, \mn@doi [\apj]
  {10.1088/0004-637X/759/1/23}, \href
  {http://adsabs.harvard.edu/abs/2012ApJ...759...23S} {759, 23}

\bibitem[\protect\citeauthoryear{{Silk} \& {Rees}}{{Silk} \&
  {Rees}}{1998}]{Silk1998}
{Silk} J.,  {Rees} M.~J.,  1998, \aap, \href
  {http://adsabs.harvard.edu/abs/1998A%26A...331L...1S} {331, L1}

\bibitem[\protect\citeauthoryear{{Spergel} et~al.,}{{Spergel}
  et~al.}{2003}]{Spergel2003}
{Spergel} D.~N.,  et~al., 2003, \mn@doi [\apjs] {10.1086/377226}, \href
  {http://adsabs.harvard.edu/abs/2003ApJS..148..175S} {148, 175}

\bibitem[\protect\citeauthoryear{{Spergel} et~al.,}{{Spergel}
  et~al.}{2007}]{Spergel2007}
{Spergel} D.~N.,  et~al., 2007, \mn@doi [\apjs] {10.1086/513700}, \href
  {http://adsabs.harvard.edu/abs/2007ApJS..170..377S} {170, 377}

\bibitem[\protect\citeauthoryear{{Stinson}, {Brook}, {Macci{\`o}}, {Wadsley},
  {Quinn}  \& {Couchman}}{{Stinson} et~al.}{2013}]{Stinson}
{Stinson} G.~S.,  {Brook} C.,  {Macci{\`o}} A.~V.,  {Wadsley} J.,  {Quinn}
  T.~R.,   {Couchman} H.~M.~P.,  2013, \mn@doi [\mnras] {10.1093/mnras/sts028},
  \href {http://adsabs.harvard.edu/abs/2013MNRAS.428..129S} {428, 129}

\bibitem[\protect\citeauthoryear{{Tanimura}, {Hinshaw}, {McCarthy}, {Van
  Waerbeke}, {Ma}, {Mead}, {Hojjati}  \& {Tr{\"o}ster}}{{Tanimura}
  et~al.}{2017}]{tSZ2}
{Tanimura} H.,  {Hinshaw} G.,  {McCarthy} I.~G.,  {Van Waerbeke} L.,  {Ma}
  Y.-Z.,  {Mead} A.,  {Hojjati} A.,   {Tr{\"o}ster} T.,  2017, preprint, \href
  {http://adsabs.harvard.edu/abs/2017arXiv170905024T} {} (\mn@eprint {arXiv}
  {1709.05024})

\bibitem[\protect\citeauthoryear{{Taylor}}{{Taylor}}{1950}]{Taylor1950}
{Taylor} G.,  1950, \mn@doi [Proceedings of the Royal Society of London Series
  A] {10.1098/rspa.1950.0049}, \href
  {http://adsabs.harvard.edu/abs/1950RSPSA.201..159T} {201, 159}

\bibitem[\protect\citeauthoryear{{Tollet} et~al.,}{{Tollet}
  et~al.}{2016}]{Tollet2016}
{Tollet} E.,  et~al., 2016, \mn@doi [\mnras] {10.1093/mnras/stv2856}, \href
  {http://adsabs.harvard.edu/abs/2016MNRAS.456.3542T} {456, 3542}

\bibitem[\protect\citeauthoryear{{Tumlinson}, {Peeples}  \& {Werk}}{{Tumlinson}
  et~al.}{2017}]{Tumlinson2017}
{Tumlinson} J.,  {Peeples} M.~S.,   {Werk} J.~K.,  2017, \mn@doi [\araa]
  {10.1146/annurev-astro-091916-055240}, \href
  {http://adsabs.harvard.edu/abs/2017ARA%26A..55..389T} {55, 389}

\bibitem[\protect\citeauthoryear{{Veilleux}, {Cecil}  \&
  {Bland-Hawthorn}}{{Veilleux} et~al.}{2005}]{Veilleux2005}
{Veilleux} S.,  {Cecil} G.,   {Bland-Hawthorn} J.,  2005, \mn@doi [\araa]
  {10.1146/annurev.astro.43.072103.150610}, \href
  {http://adsabs.harvard.edu/abs/2005ARA%26A..43..769V} {43, 769}

\bibitem[\protect\citeauthoryear{{Vogelsberger} et~al.,}{{Vogelsberger}
  et~al.}{2014}]{Vogelsberger2014}
{Vogelsberger} M.,  et~al., 2014, \mn@doi [\mnras] {10.1093/mnras/stu1536},
  \href {http://adsabs.harvard.edu/abs/2014MNRAS.444.1518V} {444, 1518}

\bibitem[\protect\citeauthoryear{{Walch} \& {Naab}}{{Walch} \&
  {Naab}}{2015}]{Walch2015}
{Walch} S.,  {Naab} T.,  2015, \mn@doi [\mnras] {10.1093/mnras/stv1155}, \href
  {http://adsabs.harvard.edu/abs/2015MNRAS.451.2757W} {451, 2757}

\bibitem[\protect\citeauthoryear{{White} \& {Frenk}}{{White} \&
  {Frenk}}{1991}]{White1991}
{White} S.~D.~M.,  {Frenk} C.~S.,  1991, \mn@doi [\apj] {10.1086/170483}, \href
  {http://adsabs.harvard.edu/abs/1991ApJ...379...52W} {379, 52}

\bibitem[\protect\citeauthoryear{{White} \& {Rees}}{{White} \&
  {Rees}}{1978}]{White1978}
{White} S.~D.~M.,  {Rees} M.~J.,  1978, \mn@doi [\mnras]
  {10.1093/mnras/183.3.341}, \href
  {http://adsabs.harvard.edu/abs/1978MNRAS.183..341W} {183, 341}

\bibitem[\protect\citeauthoryear{Yepes, Kates, Khokhlov  \& Klypin}{Yepes
  et~al.}{1997}]{Yepes}
Yepes G.,  Kates R.,  Khokhlov A.,   Klypin A.,  1997, \mn@doi [Monthly Notices
  of the Royal Astronomical Society] {10.1093/mnras/284.1.235}, 284, 235

\bibitem[\protect\citeauthoryear{{de Graaff}, {Cai}, {Heymans}  \&
  {Peacock}}{{de Graaff} et~al.}{2017}]{tSZ1}
{de Graaff} A.,  {Cai} Y.-C.,  {Heymans} C.,   {Peacock} J.~A.,  2017,
  preprint, \href {http://adsabs.harvard.edu/abs/2017arXiv170910378D} {}
  (\mn@eprint {arXiv} {1709.10378})

\makeatother
\end{thebibliography}

\appendix
\section{Analytic Expressions for the ($\alpha,\beta,\gamma$) Profile}
\label{app1}
The ($\alpha,\beta,\gamma$) profile \citep{Hernquist1990} is defined by
\begin{equation}
    \rho(r)=\frac{\rho_s}{\left(\frac{r}{r_s}\right)^{\gamma}\left(1+\left(\frac{r}{r_s}\right)^{\alpha}\right)^{\frac{\beta-\gamma}{\alpha}}},
\end{equation}
where $r_s$ and $\rho_s$ are the scale radius and scale density of the halo respectively.  At small $r$ the density has a logarithmic slope of $\gamma$, while at large $r$ it falls off as $\rho\propto r^{-\beta}$. $\alpha$ marks the transition strength between these two regimes. In general, the power-law slope of the profile can be derived as
\begin{equation}
    s(r)=-\frac{d\log\rho}{d\log r}=-\frac{\gamma+\beta\left(\frac{r}{r_s}\right)^{\alpha}}{1+\left(\frac{r}{r_s}\right)^{\alpha}}.
\end{equation}

To solve for the potential,
\begin{equation}
    \Phi(r)=-4\pi G\left(\frac{1}{r}\int^r_0\rho(r')r'^2dr'+\int^{\infty}_r\rho(r')r'dr'\right),
\end{equation}
we make the following assumptions: $\alpha>0$, $0<\gamma<2$, $\beta>2$, $\rho_s>0$, and $r_s>0$.  Evaluating these integrals we find two cases:
\begin{equation}
    \Phi(r)= 
\begin{cases}
    -4\pi G\rho_s(A_1[r]+A_2[r]),& \text{if } r> 0\\
    \frac{-4\pi G\rho_sr_s^2\Gamma\left[\frac{\beta-2}{\alpha}\right]\Gamma\left[\frac{2-\gamma}{\alpha}\right]}{\alpha\Gamma\left[\frac{\beta-\gamma}{\alpha}\right]},              & \text{if } r=0
\end{cases}
\end{equation}
where 
\begin{equation}
    A_1(r)=\frac{-r^{2-\gamma}r_s^{\gamma} \ _{2}F_1\left[\frac{3-\gamma}{\alpha},\frac{\beta-\gamma}{\alpha};\frac{3+\alpha-\gamma}{\alpha};-\left(\frac{r}{r_s}\right)^{\alpha}\right]}{\gamma-3}
\end{equation}
\begin{equation}
    A_2(r)=\frac{r^{2-\beta}r_s^{\beta} \ _{2}F_1\left[\frac{\beta-2}{\alpha},\frac{\beta-\gamma}{\alpha};\frac{\alpha+\beta-2}{\alpha};-\left(\frac{r}{r_s}\right)^{-\alpha}\right]}{\beta-2}.
\end{equation}
Here, $_2F_1[a,b;c;z]$ is the Gaussian or ordinary hypergeometric function and $\Gamma[x]$ is the gamma function.

Our assumptions on $\alpha,\ \beta,\ \&\ \gamma$ were made so that
\begin{equation}
    \lim_{r\to0}\frac{1}{r}\int^r_0\rho(r')r'^2dr'=0
\end{equation}
and
\begin{equation}
    \lim_{r\to\infty}\Phi(r)=0.
\end{equation}
These minor restrictions encompass the wide range of $\alpha,\ \beta,\ \&\ \gamma$ that real galaxies are expected to exhibit and do not limit our ability to fit rotation curves \citep{DC2014,Katz2016}.  

We can now calculate the escape speed, $V_{\rm esc}$, at any chosen radius:
\begin{equation}
    V_{\rm esc}=\sqrt{2|\Phi(r)|}.
\end{equation}
The mass distribution within the galaxy is given by
\begin{equation}
    M(r)=4\pi\rho_s\int^r_0\frac{r'^2}{\left(\frac{r'}{r_s}\right)\left(1+\left(\frac{r'}{r_s}\right)^{\alpha}\right)^{\frac{\beta-\gamma}{\alpha}}}dr'.
    \label{mor}
\end{equation}
Using our previous solution of this integral in calculating the potential:
\begin{equation}
    M(r)=4\pi\rho_s rA_1[r].
\end{equation}
This defines the circular velocity, $V_c(r)$, at any given radius by
\begin{equation}
    V_c(r)=\sqrt{4\pi G\rho_s A_1[r]}.
\end{equation}

These equations have all been given in terms of $\rho_s$ and $r_s$. We can  reformulate them in terms of concentration $c$ and virial velocity $V_{\rm vir}$ by using
\begin{equation}
    r_s=\frac{r_{-2}}{\left(\frac{2-\gamma}{\beta-2}\right)^{1/\alpha}},
    \label{rrseqn}
\end{equation}
where
\begin{equation}
    r_{-2}=\frac{R_{\rm vir}}{c}.
\end{equation}
Here, we define the concentration as $R_{\rm vir}$ divided by the radius, $r_{-2}$, where the logarithmic slope of the profile reaches the value of $-2$.  In the NFW model, this corresponds to $r_s$, in a general $(\alpha,\beta,\gamma)$ model, $r_{-2}$ is given by Equation~\ref{rrseqn}. Hence,
\begin{equation}
    r_s=\frac{R_{\rm vir}}{c\left(\frac{2-\gamma}{\beta-2}\right)^{1/\alpha}}.
\end{equation}

We can find $\rho_s$ by inverting Equation~\ref{mor} and evaluating at $R_{\rm vir}$.  For this, we have
\begin{equation}
    \rho_s=\frac{M_{\rm vir}}{4\pi R_{\rm vir}A_1[R_{\rm vir}]}.
\end{equation}
We can then place this in the context of $V_{\rm vir}$ by using
\begin{equation}
    R_{\rm vir}=\left(\frac{M_{\rm vir}}{\frac{4\pi\Delta}{3}\rho_{\rm crit}}\right)^{1/3}
\end{equation}
and
\begin{equation}
    M_{\rm vir}=\frac{V_{\rm vir}^3}{\sqrt{\frac{\Delta}{2}}GH_0},
\end{equation}
so
\begin{equation}
    R_{\rm vir}=\frac{V_{\rm vir}}{H_0\sqrt{\frac{\Delta}{2}}}.
\end{equation}
Here, $\Delta$ is the over-density with respect to the critical density, $\rho_{\rm crit}$ at which a halo is considered virialized, $G$ is the gravitational constant, and $H_0$ is the Hubble constant.

Substituting in for $A_1$ and $\rho_s$, we find
\begin{equation}
    A_1(r)=\frac{-r^{2-\gamma}\ _2F_1\left[\frac{3-\gamma}{\alpha},\frac{\beta-\gamma}{\alpha};\frac{3+\alpha-\gamma}{\alpha};-\left(\frac{r\sqrt{\frac{\Delta}{2}}H_0c}{V_{\rm vir}}\right)^{\alpha}\right]}{\left(\frac{V_{\rm vir}}{\sqrt{\frac{\Delta}{2}}H_0c}\right)^{-\gamma}(\gamma-3)}
\end{equation}
and
\begin{equation}
    \rho_s=\frac{V_{\rm vir}^2}{4\pi G A_1\left[\frac{V_{\rm vir}}{\sqrt{\frac{\Delta}{2}}H_0}\right]}.
\end{equation}
We can now calculate $\Phi$, $M(r)$, and $V_c(r)$ solely from $c$ and $V_{\rm vir}$ without the use of numerical integrals.

In this work, when we consider the ejection model, we will assume that all baryons need to be ejected from the centre of the halo.  While it is unlikely that all of the gas is at $r=0$, stars tend to form at the centres of galaxies in the highest density gas, so it is likely that the outflowing gas originates at $r\ll{R_{\rm vir}}$.  While this may be overestimating the true ${V_{\rm esc}}$ of a small parcel of gas, we will assume later that the rest of the baryonic matter is evenly spread throughout the halo, which causes us to underestimate ${V_{\rm esc}}$ from the centre. Although simple, our assumptions are therefore not obviously systematically biased.

We now have the means to use an analytic set of functions to calculate ${V_{\rm esc}}$ as a function of $c$, ${V_{\rm vir}}$, $\alpha$, $\beta$, and $\gamma$.  Putting this all together, we find
\begin{equation}
\begin{aligned}
    {V_{\rm esc}(r=0)}=\\
    \sqrt{2\left|\frac{(\gamma-3){V_{\rm vir}^2}\left(\frac{1-\gamma}{\beta-2}\right)^{(2-\gamma)/\alpha}\Gamma\left(\frac{\beta-2}{\alpha}\right)\Gamma\left(\frac{2-\gamma}{\alpha}\right)}{\alpha c^{2-\gamma}\Gamma\left(\frac{\beta-\gamma}{\alpha}\right){\rm _2F_1}\left[\frac{3-\gamma}{\alpha},\frac{\beta-\gamma}{\alpha};\frac{3+\alpha-\gamma}{\alpha};-c\left(\frac{1-\gamma}{\beta-2}\right)\right]}\right|}.
\end{aligned}
\end{equation}

For the NFW halo ($\alpha=1,\ \beta=3,\ \gamma=1$), these equations simplify considerably.  In this special case, we find that 
\begin{equation}
    \Phi_{\rm NFW}(r=0)=\frac{-c{V_{\rm vir}^2}}{\log(1+c)-\frac{c}{1+c}}.
\end{equation}
Therefore
\begin{equation}
    {V_{\rm esc,NFW}}=\sqrt{2\left|\frac{-c{V_{\rm vir}^2}}{\log(1+c)-\frac{c}{1+c}}\right|}.
\end{equation}

\begin{figure}
\centerline{\includegraphics[scale=1]{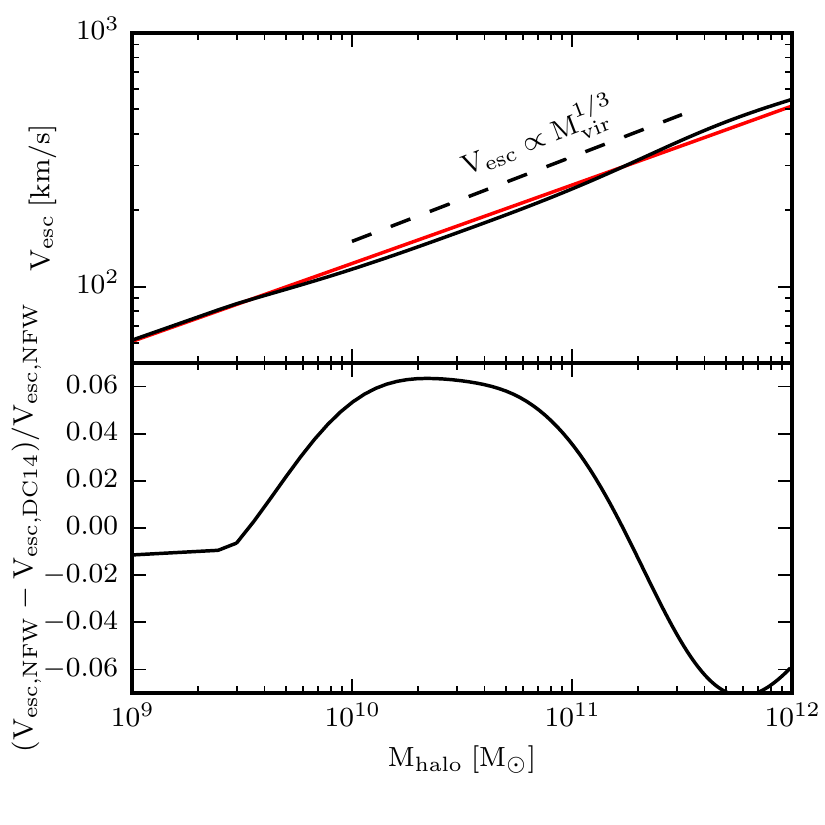}}
\caption{ (Top) Comparison of the escape velocity of an NFW halo (red) versus a DC14 halo (black) with properties that fall on the ${\rm M_*/M_{halo}}$ relation as given in \protect\cite{Moster2013} and the ${\rm M_{vir}}-c$ relation as given by \protect\cite{Dutton2014}, as a function of halo mass.  (Bottom) Percentage difference between the NFW and DC14 halo models.}
\label{vesccomp}
\end{figure}

To demonstrate the difference between the two halo models, in Figure~\ref{vesccomp} we compare ${V_{\rm esc}}$ for the NFW and DC14 models as a function of halo mass for galaxies that fall on the ${\rm M_*/M_{halo}}$ relation as given in \cite{Moster2013} and the ${\rm M_{vir}}-c$ relation as given by \cite{Dutton2014}. The maximum difference in ${V_{\rm esc}}$ for these galaxies is $\sim7\%$. Since we expect galaxies to fall close to these relations, it is reasonable to assert that for both of these two halo models, ${V_{\rm esc}\propto V_{\rm vir}\propto {\rm M_{vir}}^{1/3}}$. This scaling can be seen in Figure~\ref{vesccomp}.

\end{document}